\documentclass[useAMS,usenatbib]{mn2e}
\usepackage{graphicx}
\usepackage{amssymb}
\usepackage{natbib}
\usepackage{verbatim}
\usepackage{booktabs}

\newcommand{\stdFig}[4]{
\begin{figure}
\center
\includegraphics[scale=#4]{#1}
\caption{\small  #2 }
\label{#3}
\end{figure}
}

\newcommand{\stdFullFig}[4]{
\begin{figure*}
\center
\includegraphics[scale=#4]{#1}
\caption{\small  #2 }
\label{#3}
\end{figure*}
}

\newcommand{\dubFig}[5]{
\begin{figure*}
\center
\includegraphics[scale=#5]{#1}
\includegraphics[scale=#5]{#2}
\caption{\small  #3 }
\label{#4}
\end{figure*}
}

\newcommand{\stdTable}[4]{
\begin{table}
\label{#3}
\begin{center}
\bgroup 
\caption{#1}
\begin{small}
\renewcommand{\tabcolsep}{12pt}
\vspace{5 pt}
#2
\end{small}
\egroup
\end{center}
\end{table}
}

\newcommand{\stdFullTable}[4]{
\begin{table*}
\begin{center}
\bgroup 
\caption{#1}
\begin{small}
\renewcommand{\tabcolsep}{12pt}
\vspace{5 pt}
#2
\end{small}
\egroup
\end{center}
\label{#3}
\end{table*}
}

\newcommand{\Alfven}{Alfv\'{e}n }
\newcommand{\msun}{\;{\rm M}_{\odot}}

\begin{document}

\title{Cosmic-ray Driven Outflows in Global Galaxy Disk Models}

\author[Salem \& Bryan]{Munier Salem$^{1}$, Greg L. Bryan$^{1}$ \\
$^{1}$Department of Astronomy, Columbia University, 550 West 120th Street, New York, NY 10027, USA}

\date{}

\maketitle



\begin{abstract}
Galactic-scale winds are a generic feature of massive galaxies with high star formation rates across a broad range of redshifts. Despite their importance, a detailed physical understanding of what drives these mass-loaded global flows has remained elusive. In this paper, we explore the dynamical impact of cosmic rays by performing the first three-dimensional, adaptive mesh refinement simulations of an isolated starbursting galaxy that includes a basic model for the production, dynamics and diffusion of galactic cosmic rays.  We find that including cosmic rays naturally leads to robust, massive, bipolar outflows from our $10^{12}M_\odot$ halo, with a mass-loading factor $\dot{M}/{\rm SFR} = 0.3$ for our fiducial run. Other reasonable parameter choices led to mass-loading factors above unity.  The wind is multiphase and is accelerated to velocities well in excess of the escape velocity.  We employ a two-fluid model for the thermal gas and relativistic CR plasma and model a range of physics relevant to galaxy formation, including radiative cooling, shocks, self-gravity, star formation, supernovae feedback into both the thermal and CR gas, and isotropic CR diffusion. Injecting cosmic rays into star-forming regions can provide significant pressure support for the interstellar medium, suppressing star formation and thickening the disk. We find that CR diffusion plays a central role in driving superwinds, rapidly transferring long-lived CRs from the highest density regions of the disk to the ISM at large, where their pressure gradient can smoothly accelerate the gas out of the disk.
\end{abstract}


\section{Introduction}

Galaxy formation, in the broadest strokes, is a story of cooling, collapse, and infall, as baryons settle into dark matter halos and form stars \citep[e.g.,][]{Rees1977, White1978}. But a more precise account of this process reveals an important role for heating, expansion, and outflow, as stars and black holes release energy, momentum and material back into the interstellar medium (ISM) and beyond, strongly impacting the resulting structure. Galactic-scale winds are among the most important of these latter processes, as they can remove appreciable mass from dense, star-forming regions. This feedback can substantially alter the distribution of luminous matter: studies that match the observed bright galaxies with their required dark matter halos find that typically only 20\% of the baryons are accounted for in $L_*$ galaxies, with the fraction decreasing for both larger and smaller systems \citep{Vale2004, Conroy2006, Behroozi2010, Guo2010}.  These estimates also agree reasonably well with the dark matter content of individual galaxies estimated either with rotation curves \citep[e.g.,][]{McGaugh2000, Stark2009, McGaugh2010} or weak lensing \citep[e.g.,][]{Mandelbaum2006}.  Generally, simulations have predicted much higher baryon fractions, producing an offset in the Tully-Fisher relation \citep[e.g.,][]{Steinmetz2002}, although very strong feedback appears to be capable of solving this issue \citep[e.g.,][]{Brook2012}. Beyond aiding in this mass displacement, galactic winds also carry energy and metals beyond their host halo, polluting the intergalactic medium \citep[e.g.,][]{Cowie1995, Porciani2005}.

Most actively star-forming galaxies with high specific star-formation rates (specific SFRs) host galactic-scale outflows \citep[see][for a recent review of galactic winds and an account of these observations]{Veilleux2005}. Both locally and at high redshift, these highly productive systems can often direct more mass into the outflowing winds than into newly formed stars, i.e. their mass loading factor (the ratio of of mass loss from the system to the SFR) is above unity \citep{Martin1999,Steidel2010}. Multiple gas phases comprise these flows, with pockets of neutral, warm-ionized and soft X-ray gas observed traveling at hundreds of km/s relative to their host galaxies \citep{Heckman1990,Pettini2001,Chen2010,Rubin2010}. 

Despite the importance and ubiquity of galactic winds, the driving mechanisms are not well understood. Many models have assumed hot evacuated gas from repeated SN drives the wind \citep{Larson1974,Chevalier1985,Dekel1986}, though more detailed simulations able to resolve interacting SN have failed to produce large mass-loading, particularly for gas-rich disks \citep{MacLow1999, Joung2009, Creasey2013}. These results suggest an important role for less obvious processes, including radiation pressure \citep[e.g.,][]{Murray2005, Nath2009, Murray2011} and the ISM's relativistic plasma component --- i.e. cosmic rays 
\citep{Boulares1990,Breitschwerdt1993,Uhlig2012}. This paper reports work exploring this final mechanism.

Though long established as an important component of the ISM's pressure support, the dynamical impact of cosmic rays of galactic winds (CRs) has received relatively little attention in recent years.  Chemical abundances of CRs tell us their lifetime in the Galactic disk is $\sim3$ Myr, although rays may enter the halo and re\"{e}nter the disk, leading to an overall estimate of $\sim20$ Myr \citep{Shapiro1970, Kulsrud2005}. Despite this short lifetime, their energy density remains significant since they are thought to be produced via shock acceleration in supernovae (SN) \citep[e.g.,][]{Blandford1987}. 

In the solar neighborhood, CRs have an observed energy density of $\sim10^{-12}$ ergs cm$^{-3}$ \citep{Wefel1987}, making CRs at least as important to model as turbulent pressure and magnetic fields.  Observations of gamma rays indicate that star-bursting galaxies have even higher CR energy densities than the Milky Way \citep{Acciari2009, Paglione2012}.  The cosmic ray pressure is contributed largely by GeV protons which scatter off of inhomogeneties in the magnetic fields, leading to the observed isotropic distribution despite the rarity of SN, their presumed acceleration sites.  CRs stream along their own pressure gradient, but if the streaming occurs faster than the \Alfven speed, they excite waves in the magnetic field which can damp and heat the gas\citep{Cesarsky1980,Wentzel1969,Kulsrud1969,Kulsrud1971}. 

It has often been argued that this additional pressure term is unimportant to the dynamics since CR diffusion ought to rapidly wash out pressure gradients that could drive gas flows. However, models suggest high star formation rates can produce and maintain a local enhancement of cosmic ray pressure capable of driving large-scale winds and regulating star formation \citep[e.g.][]{McKenzie1987, Breitschwerdt1991, Socrates2008, Dorfi2013}. In addition, the role of diffusion diminishes on larger length scales, suggesting that CRs may be important for suppressing large-scale disk fragmentation (which is not observed), while still allowing smaller scale molecular clouds to form.

Previous work on CR dynamics, while rare, has investigated many important issues.  \cite{Breitschwerdt1991} used a 1D time-independent model to explore how CRs and MHD effects may help drive these outflows. In their picture, the CRs stream along a large-scale, coherent magnetic field's ``flux tubes'', oriented vertically close to the disk and radially far from the galaxy. The rays exert a pressure on the thermal gas via scattering with the field and an \Alfven wave pressure is also explicitly included. This model supported an important role for CRs in accelerating thermal gas beyond the disk plane, gently at first, but increasing to high speeds far above the star forming disk. Subsequent work has enhanced this simple model to include effects in the disk-halo interface, slightly modified magnetic field geometries, disk rotation/magnetic tension, and more sophisticated damping mechanisms~\citep{Breitschwerdt1993,Zirakashvili1996,Everett2008}. Recently, \cite{Dorfi2013} further extended these models to include time-dependent outflows and CR diffusion, still in a 1D setting. This enhanced approach found bursty, transient winds featuring forward- and backward-propagating  shock fronts, with implications for diffuse shock acceleration of CRs. These 1D models make a persuasive case for the dynamical importance of CRs but can only treat aspects of the flow inherent to a uniform, coherent wind along ordered field lines. In reality, the dynamics are likely to involve multiphase, turbulent gas flows and field lines with far richer topologies. In addition, all these models treat the flow of gas and rays beyond the disk as an inner boundary condition, and do not explore how gas and rays are produced within and rise out of the patchwork star forming regions of a real disk.

CRs have only recently been incorporated into 3D, global hydrodynamic simulations for preliminary explorations into their dynamical role in galaxy evolution. \cite{Ensslin2007} and \cite{Jubelgas2008} modified the SPH code Gadget to include CRs and used it to examine both idealized and cosmological simulations of galaxies. They found CRs can significantly suppress the star formation rates and other properties of galaxies with circular velocities less than 80 km s$^{-1}$. These pioneering studies demonstrated the importance of CRs, but had a number of shortcomings: most production runs did not include CR diffusion or streaming; most cosmological runs were at high redshift, and the simulations included a `stiff' thermal equation of state from the \cite{SnH2003} subgrid model, which already builds in feedback to suppress disk fragmentation. Very recently, \cite{Uhlig2012} built on this earlier work by including cosmic-ray streaming and found the production of significant outflows, although again the simulations did not include CR diffusion, used the \cite{SnH2003} subgrid model, and were not cosmological. No simulation at the galactic scale has explicitly included magnetic fields or dealt with the anisotropies they may introduce.

In this paper, we adopt a simple two-fluid model for the cosmic rays and thermal plasma to explore the dynamical impact of the CR pressure on high resolution, global simulations of an idealized $10^{12}M_\odot$ disk galaxy.  We assume the rays can be treated as a relativistic plasma of negligible inertia that is tied to the thermal plasma except for an isotropic diffusion term.  We also include source terms for the CRs under the assumption that they are mostly produced in strong shock waves generated by supernovae.  As we will show, this model -- although simple -- can drive significant outflows.  In Section~\ref{sec:methodology} we describe the CR model and initial conditions; in Section~\ref{sec:results}, we describe the results of our numerical experiments, and finally, in Section~\ref{sec:discussion}, we describe a simple picture to understand our results, and discuss both implications and shortcomings of this work.

\section{Methodology}
\label{sec:methodology}

\subsection{The Two Fluid Model for Gas and CRs}

We begin with a two-fluid approach to modeling cosmic rays \citep{Jun1994,Drury1985,
Drury1986}. The model assumes an ultra-relativistic gas of protons which we treat as an ideal
gas with $\gamma = 4/3$ that is tied to the thermal plasma except for a diffusion term. 
While a detailed treatment of the high energy
particles involving both a distribution in momentum space and a treatment of magnetic fields
could lead to an anisotropic CR pressure on the gas, the two-fluid approach assumes a 
scalar pressure to make the model tractable. Observations are consistent with an isotropic
distribution of particles, particularly in the GeV range which contributes primarily to the pressure.
We neglect diffusion of the cosmic rays in energy, as well as energy loss terms due to direct
collisions or to interactions with the magnetic field. We further assume the large-scale 
magnetic field to be dynamically sub-dominant.
These assumptions lead to the following set of equations \citep{Drury1985}
\begin{eqnarray}
\label{mass}
\partial_t \rho + \nabla \cdot  ( \rho \bf{u} ) 					&=& 	0 ,  \\ 
\label{momentum}
\rho \left( \partial_t \bf{u} +  \bf{u} \cdot \nabla \bf{u} \right) 		&=&	- \nabla ( P_{\rm th} + P_{\rm cr} ) , \\
\label{energy}
\partial_t \epsilon_{\rm th} + \nabla \cdot ( \epsilon_{\rm th} \bf{u} )	&=&	- P_{\rm th} ( \nabla \cdot \bf{u} ) + \Gamma + \Lambda , \\
\label{cr_energy}
\partial_t \epsilon_{\rm cr} + \nabla \cdot ( \epsilon_{\rm cr} \bf{u} )	&=&	 - P_{\rm cr} ( \nabla \cdot \bf{u} ) \\
									& &	+ \nabla \cdot ( \kappa_{\rm cr} \nabla \epsilon_{\rm cr} ) + \Gamma_{\rm CR} , 
\end{eqnarray}
along with the state equations 
\begin{eqnarray}
\label{gas_state}
P_{\rm th} &=& ( \gamma_{\rm th} - 1 ) \epsilon_{\rm th} , \\
\label{cr_state}
P_{\rm cr} &=& ( \gamma_{\rm cr} - 1 ) \epsilon_{\rm cr} ,
\end{eqnarray}
where $\rho$ is the gas density, $\bf{u}$ is the gas velocity, $P$ is the pressure, $\epsilon$ is the energy density and $\gamma$ is the adiabatic index. The constant $\kappa_{cr}$ is the CR diffusion coefficient, which we treat as isotropic and independent of any of our state quantities. $\Gamma$ and $\Lambda$ represent source and loss terms for the fluid. In our galaxy simulations both the CR and thermal fluids receive energy injections within star-forming regions. In these runs the thermal gas is also subject to radiative cooling. We ignore CR loss terms in our present work.

These equations represent the standard Euler equations of an ideal fluid, with a second (diffusive) CR fluid that interacts with the gas only via the momentum equation. Note that the cosmic ray mass density is negligible, allowing us to ignore it in Equation \ref{mass}. Here we implicitly ignore the motion of scatterers relative to the fluid, while still accounting for this process as diffusion of the CR energy density, $\epsilon_{cr}$.  We will discuss some possible ramifications of these assumptions later in this paper; however, we are interested in first exploring a simple model that both captures the key effect and allows us to carry out a relatively large number of simulations.

\subsection{Implementation}
\label{sec:implementation}

Our CR model was integrated into the well-tested Eulerian hydrodynamics code Enzo, described in~\cite{Bryan2013}, ~\cite{Bryan1997},~\cite{Bryan1999},~\cite{Norman1999},~\cite{Bryan2001} and~\cite{OShea2004}. One of Enzo's main strengths is adaptive mesh refinement (AMR), which uniformly resolves the entire simulation region on a course grid but provides higher resolution, ``refined'' sub-grids as needed in regions where the dynamics grow complex.

With our new two-fluid model, the list of conserved quantities in the code grows to include the CR energy density, $\epsilon_{\rm cr}$. For our preliminary investigations, we opted to work with the simple and robust ZEUS hydro method~\citep{Stone1992}, where the equations are broken into source and transport steps. The CR modifications leave the transport step unaltered. Passing this new quantity to the transport solver automatically implements the right-hand side of Equation \ref{cr_energy}, where the new CR energy density is advected with the thermal fluid.

Next, within the source step, we have implemented the pressure gradient term in Equation \ref{momentum} and the the first left-hand side term in Equation \ref{cr_energy}. These terms represent work done by the rays on the thermal gas, and the loss in CR energy density due to that work, respectively. As in the case of the hydro quantities, simple, explicit, centered-difference derivatives were used
\footnote{Following the original ZEUS implementation, Enzo stores vector quantities on cell faces, and scalar quantities at the center of cells. Thus the spatial components of material derivatives are automatically centered-difference (second order).}.

\stdFullFig{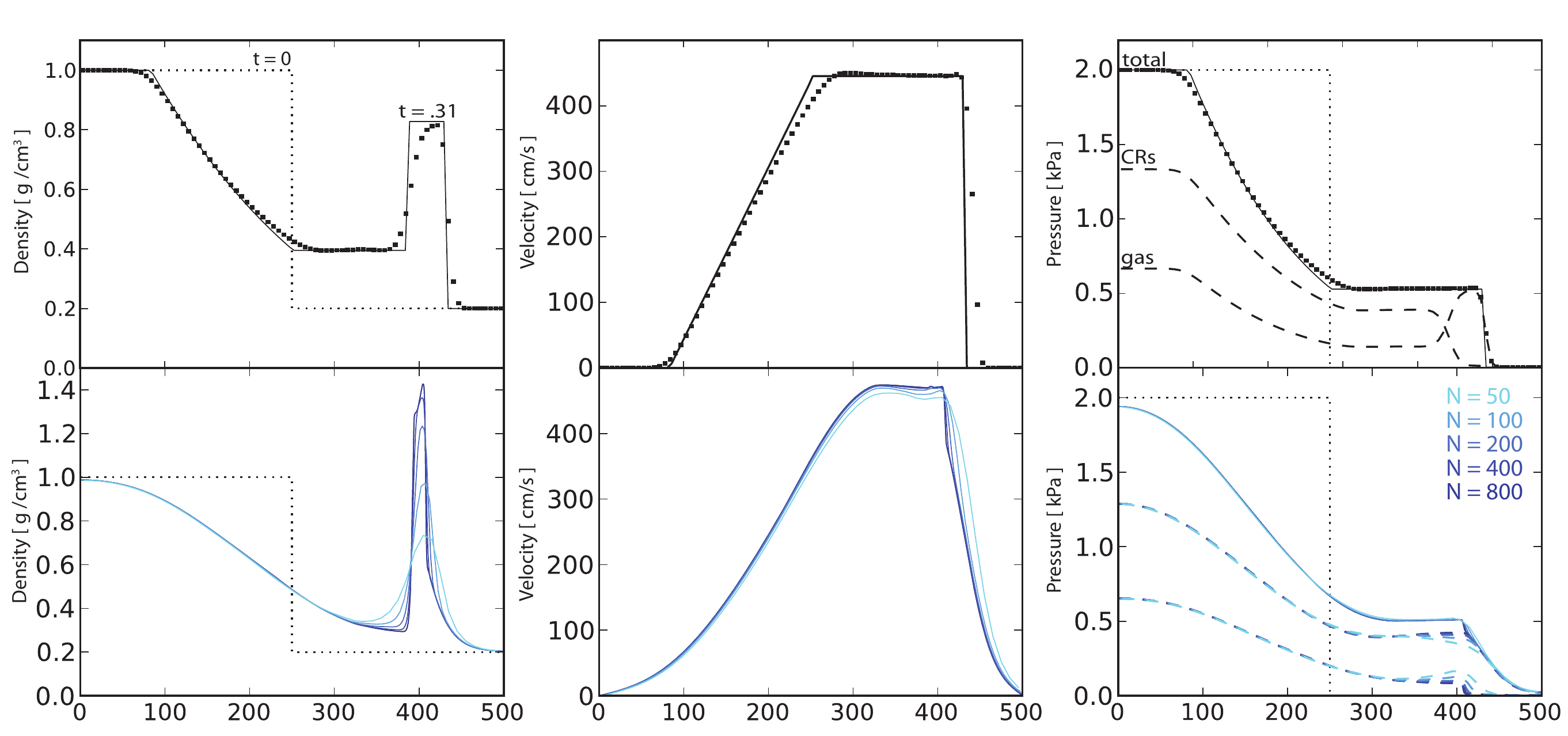}{The top row shows the CR-modified shock tube problem of P06, plotting (from left to right), the density, velocity and pressure (both CR and thermal).  Dotted lines show the $t=0$ initial conditions, while solid lines indicate the analytic result.  No CR diffusion is used.  The bottom row shows the same quantities for a run \emph{with diffusion}, where $\kappa_{\rm cr} = 0.1$. The line shade indicates the resolution used, from 50 cells (light blue) to 800 cells (dark blue).}{shocktube}{.65}

The fastest information can propagate across a fluid is the sound speed (important for subsonic flows) compounded with the bulk speed (important for super-sonic flows). For accuracy and stability, our time step must remain smaller than the time it takes information to cross an entire grid cell. In a standard fluid, the sound speed can be derived from the thermal pressure. For our new two-fluid code, we may define an effective sound speed,
\begin{equation}
\label{effectiveSoundSpeed}
c_{\rm eff} \equiv \sqrt{ \frac{\gamma_{\rm eff} P_{\rm T}}{\rho} } ,
\end{equation}
where the total pressure $P_{\rm T} = P_{\rm th} + P_{\rm cr}$, and $\gamma_{\rm eff}
 \equiv \left( \gamma_{\rm th} P_{\rm th} + \gamma_{\rm cr} P_{\rm cr} \right) / P_{\rm T}$.
In practice, we found small, low density, CR-dominated cavities can develop above our disk during bursts of star formation. Within these pockets, the effective sound speed becomes enormous since
\begin{equation}
c_{\rm s, eff} \equiv \sqrt{\gamma_{\rm eff}\frac{P_{\rm T}}{\rho}} \approx \sqrt{\frac{\gamma_{\rm cr}P_{\rm cr}}{\rho}}
\end{equation}
where $P_{\rm cr}/P_{\rm th} >> 1$. A high sound speed within a low density pocket can cause our computations to grind to a halt, since the 
time step will be limited by the Courant condition
\begin{equation}
\Delta t \propto \frac{\Delta x}{c_{\rm s, eff} + u} \approx \Delta x \frac{1}{c_{\rm s, eff}} \propto \Delta x \left( \frac{\rho}{P_{\rm CR}} \right)^{1/2}
\end{equation}
where $u$ is the magnitude of the fluid speed, which is appreciably smaller than $c_{\rm s,eff}$ in the regions of interest. This relationship suggests that raising the density within these regions can speed the pace of our runs, and since these regions are very much CR-dominated the artificially enhanced density should not substantially change the dynamics. We place an upper limit on the allowed effective sound speed by increasing the gas density in cells that exceed this limit so that $c_{\rm s,eff} < c_{\rm s,max}$. We explore the implications of this artificial ceiling in our parameter study below, where we find the choice does not substantially affect our results.

The diffusive term in Equation \ref{cr_energy} is likewise implemented with an explicit finite-difference scheme. To ensure stability, the time step of our diffusion scheme should remain smaller than the time it takes information to propagate beyond our differencing scheme's domain of dependence. To ensure this, multiple time steps may be taken within our CR-diffusion scheme for every source time step, no larger than
\begin{equation}
\Delta t_{\rm CRdiffusion} < \frac{1}{2N} \frac{\Delta x^2}{\kappa_{\rm cr}},
\end{equation}
where $N$ is the dimensionality of our simulation. This sub-cycling is limited by the number of ghost cells buffering each sub-grid, since otherwise the fluid would diffuse beyond a sub-grid before its parent grid could be made aware.

%
\subsection{Two-Fluid Model Tests}
\stdTable{Riemann Problem Parameters}{
	\begin{tabular}{lll}
		\toprule		
						& Post-Shock			& Pre Shock		\\
		\midrule
		$\rho$			& 1.0				& 0.2			\\
		$P_{\rm th}$		& $2/3 \times 10^5$	& 267.2			\\
		$\epsilon_{\rm cr}$	& $4.0 \times 10^5$	& 801.6			\\
		$v$				& 0.0				& 0.0			\\
		\bottomrule
	\end{tabular}
}{shockValues}

The two-fluid model admits an analytic solution to the Riemann problem for non-diffusive CRs ($\kappa_{\rm CR} = 0.0$), an extension of the classic Sod Shock-tube problem described in \cite{Sod1978}. The analytic solution for the two-fluid case is derived in \cite{Pfrommer2006} (hereafter P06). The resulting evolution is qualitatively similar to the classic case: A shock front and contact discontinuity (CD) propagate forward and a rarefaction fan spreads back as characteristics send word of the initial disequilibrium through the domain. The pressure and density profiles are significantly modified, as the disparity in $\epsilon_{\rm CR}$ causes a jump in thermal pressure at the CD (where the total pressure remains identical on either side for all time). This leads to a significant enhancement of the density between the shock front and CD. 

We reproduce this result in a 1D Enzo simulation using our Zeus hydro scheme. Our initial conditions, described in table \ref{shockValues} were chosen to produce a shock with Mach Number, $\mathcal{M} = 10.0$. We follow P06 and define an effective Mach number
\begin{equation}
\label{mach}
\mathcal{M} \equiv \sqrt{ \frac{\left( P_{\rm T, 1} - P_{T, 5}\right)x_s}{\rho_5c_{\rm eff, 5}^2\left( x_s - 1 \right)} } ,
\end{equation}
where Region 5 is the low-density, pre-shock region (right) and Region 1 is the high density post-shock region. Regions 2 -- 4 appear as the system evolves, as described in P06 $x_s = \rho_1/\rho_5$ and $c_{\rm eff}$ and $\gamma_{\rm eff}$ are given by Equation \ref{effectiveSoundSpeed}.

The top row of Figure \ref{shocktube} shows the results for a modest resolution of $N = 80$. At higher resolution or with an adaptive mesh the solution converges nicely to the analytic case, largely devoid of any spurious oscillations or overshoot. As seen here, low resolution runs can be quite diffusive, smearing out the result at non-smooth points in the solution, particularly at the CD. Enzo also agreed with this solution when we ran the problem in 3D with plane-parallel initial conditions. 

\stdFig{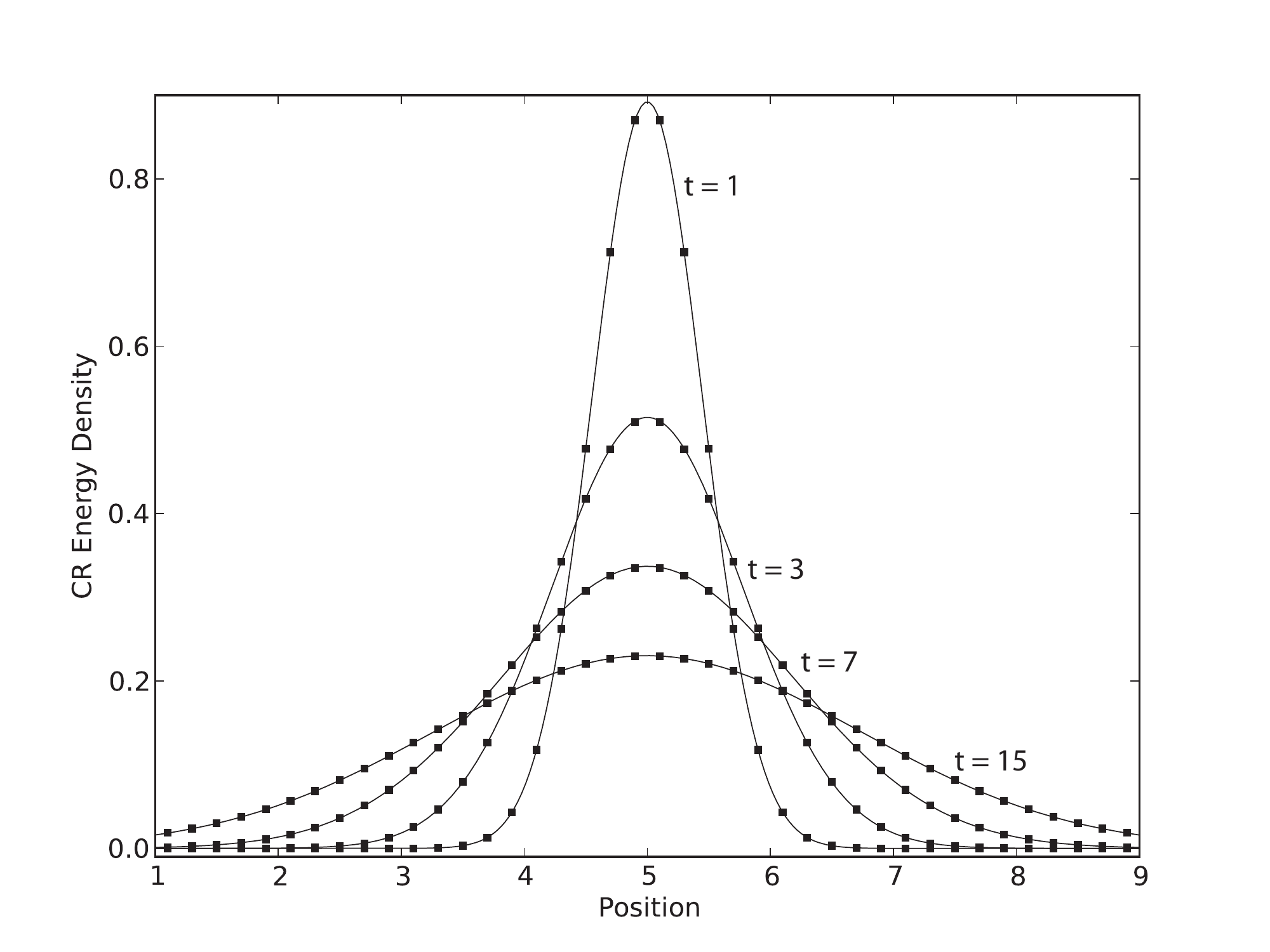}{The CR energy density as a function of time for our simple diffusion test (squares).  The analytic solution is shown as solid lines.}{diffusion}{.45}

Diffusion of the CR fluid will prove central to our investigation of CR-driven outflows. The bottom row of Figure \ref{shocktube} shows a solution to the same Riemann problem, but now for $\kappa_{\rm CR} = 0.1$. Convergence at high resolution and the absence of any spurious oscillations leads us to conclude the diffusion scheme is stable, even in the presence of shocks, CDs, sonic points and local extrema in both fluid quantities. Beyond the obvious effects of diffusion, a noticeable spike in gas density occurs behind the shock front.  This density spike is a classic feature of diffusive shock acceleration \citep[e.g.,][]{Jun1994, Jun1997}. 

We can test the diffusion scheme by itself in a more quantitative manner for a simpler test problem: We fill a one-dimensional domain with high density gas ($\rho = 10,000$ in code units) at rest in a region of uniform pressure. We then place a small amplitude, local enhancement of CRs at the center of this domain. The thermal gas here has too much inertia to be altered by the CR pressure over the relevant diffusion timescale. Thus the two-fluid model reduces to a simple, linear diffusion equation for constant $\kappa_{\rm CR}$: $\partial_t \epsilon_{\rm cr} = \kappa_{\rm cr} \partial_x^2 \epsilon_{\rm cr}$. This
equation admits a classic analytic solution for $\epsilon_{\rm cr}(t=0) = \delta(x-x_0)$, given by
\begin{equation}
\epsilon_{\rm cr}(x,t) = \frac{1}{\sqrt{4\pi \kappa_{\rm CR}}} \exp \left( - \frac{x^2}{4 \kappa_{\rm CR} t} \right) \; .
\end{equation}
We evolve this solution, beginning at $t=1.0$ for the case where $\kappa_{\rm CR} = 0.1$, as shown in Figure \ref{diffusion}. The simple finite difference scheme does a great job matching the solution, here for $N = 50$ grid cells.


\stdTable{Galaxy Parameters}{
\begin{tabular}{ll}
\toprule		
\textbf{NFW Profile}		& 	\\
$M_{200}$			& $1.0 \times 10^{12} M_\odot$	\\
$c$					& $12.0$						\\
\midrule
\textbf{Disk}			&							\\
$M_{\rm disk}$			& $6.0 \times 10^{10} $ $M_\odot$	\\
$f_{\rm gas,0}$			& $1.0$						\\
$H$					& $.325$ kpc					\\
$R$					& $3.5$ kpc					\\
\bottomrule
\end{tabular}
}{tab:galParams}

\subsection{Initial Conditions}

Our work follows that of \citet{Tasker2006}, whose ICs, summarized in Table 2, provided the point of departure for our simulations. These runs include radiative cooling of the thermal gas but no cooling of the CR component (see \ref{sec:missingPhysics}).

Our runs begin with an isothermal gas disk at $10^4$ K whose density follows
\begin{equation}
\rho_{\rm D}(r,z) = \rho_0 e^{- \frac{r}{r_0}} {\rm sech}^2 \left( \frac{1}{2} \frac{z}{z_0} \right). 
\end{equation}
where we set the vertical scale height to $z_0 = 325$ pc and the radial scale height to $r_0 =  3.5$ kpc. We set the 
total disk gas mass to $ 6 \times 10^{10} \msun$ which gives us a $\rho_0 \sim 10^{-20}$ kg m$^{-3}$. This total 
mass is roughly that of the Milky Way total disk components (stars and gas -- e.g.~\cite{Klypin2002}).

In addition to the disk's self gravity, it sits in a static dark matter potential with the standard form \citep{Navarro1997}, given (in spherical 
coordinates) by
\begin{equation}
M_{\rm DM}(R) = \frac{M_{200}}{f(c)} \left[ \ln\left( 1 + x \right) - \frac{x}{1 + x } \right].
\end{equation}
We set the virial mass, $M_{200}$, to $10^{12} \msun$. The dimensionless radius $x = Rc/r_{200}$ where c is the 
concentration parameter, which we set to $c = 12$. $f(c)$ is given by
\begin{equation}
f(c) = \ln \left( 1 + c \right) - \frac{c}{1+c} .
\end{equation}
To begin our runs in mechanical equilibrium given our gaseous and dark matter mass, $M_{\rm tot}$, we set the orbital 
speed within the disk to $V_{\rm circ}(R) = (GM_{\rm tot}/R)^{1/2}$.

For runs including CRs, we begin by adopting a simple prescription that maps CR energy density, $\epsilon_{\rm cr}$, to 
gas density, $\rho$, by
\begin{equation}
\label{crudeCR}
\epsilon_{\rm cr}(r,z) = \alpha_{\rm cr} \times \rho(r,z)
\end{equation}
in dimensionless code units. For standard runs, we set $\alpha_{\rm cr} = 0.1$, which corresponds to $\epsilon_{\rm cr} 
\approx 3 \times 10^{-12}$ ergs/cm$^3$ in the solar neighborhood, in line with laboratory results. Although this setup is
not realistic, the generation and diffusion of CR rays quickly dominates the CR distribution and
the choice of CR initial conditions has only a tiny effect on our results.

Our galaxy sits at the center of a $(500$ kpc$)^3$ box, partitioned into $128^3$ cells. Within regions where density exceeds the background density by a factor of four, enzo instantiates a higher resolution ``sub-grid'', rebuilt at each time step, that increases resolution by a factor of two. This refinement occurs recursively, and we allow up to six levels of refinement in our fiducial run, for an effective spatial resolution of $61$ pc in the highest density regions (the majority of the galactic  disk). 

\stdFullTable{Varied Parameters}{
\begin{tabular}{lcccccccc}
\toprule		
\hline
		& $\Delta x_{\rm min}$*	& size*	& $c_{s, {\rm max}}$& $\epsilon_{\rm SF}$	& $\epsilon_{\rm SN}$	& $f_{\rm CR}$& $\kappa_{\rm CR}$	& $\gamma_{\rm CR}$	\\
\midrule
Very High	& ---					& ---		& ---				& ---					& ---					& ---			& $1\times10^{29}$	& $5/3$	\\
High		& 31.	 pc				& $256^3$& 10,550	km/s		& .05					& $3\times10^{-6}$		& 1.0			& $3\times10^{28}$	& $3/2$	\\
Fiducial	& 61.	 pc				& $128^3$& 3,518 km/s		& .01					& $1\times10^{-5}$		& 0.3			& $1\times10^{28}$	& $4/3$	\\
Low		& 122. pc				& ---		& 1,550 km/s		& ---					& $3\times10^{-6}$		& 0.0			& $3\times10^{27}$	& ---		\\
Very Low	& ---					&---		& ---				& ---					& ---					& ---			& 0				& ---		\\
\bottomrule
\label{tab:params}
\end{tabular}
}{parameters}

\subsection{Star Formation and Feedback}

For star formation, we follow the prescription of \cite{Cen1992}, with updates first described in \cite{OShea2004}. A cell in Enzo 
will produce a ``star particle'' if: (1) the gas density exceeds a threshold density; (2) the gas mass of the cell exceeds the local 
Jeans mass; (3) the flow converges, i.e. $\nabla \cdot \textbf{v} < 0$; and (4) the dynamical time exceeds the gas cooling time, 
$t_{\rm cool} < t_{\rm dyn}$, or the temperature is at the minimum allowed value. Pursuant to these conditions, Enzo siphons 
gas from the grid cell into a star particle of mass
\begin{equation}
m_\star = \epsilon_{\rm SF} \frac{\Delta t}{t_{\rm dyn}} \rho_{\rm gas} \Delta x^3,
\label{starParticleMass}
\end{equation}
where $\epsilon_{\rm SF}$ is the the star formation efficiency. \cite{Tasker2006} found that $\epsilon_{\rm SF} = .05$ does a good
job reproducing the global Schmidt-Kennicut law, and we adopt this value for our fiducial run. To prevent an excess of small star 
particles bogging down our computation, we set a minimum $m_{\star, \rm min} = 10^5 M_\odot$. For cells where 
$m_\star < m_{\star, \rm min}$  is the only obstacle to forming a star particle, a particle may still be created with a probability 
$m_\star/m_{\star, \rm min}$ whose mass is either the minimum mass or 80\% of the cell mass, whichever is smaller. The particle's 
creation occurs over a dynamical time, its mass grows following
\begin{equation}
m_\star(t) = m_\star \int_{t_0}^t \frac{t - t_0}{\tau^2}\exp\left( -\frac{t - t_0}{\tau}\right) dt
\end{equation}
where $m_\star$ on the right hand side is the final mass of the particle from Equation \ref{starParticleMass}, $t_0$ is when 
the particle's formation began, and $\tau = {\rm max}(t_{\rm dyn},10 \;{\rm Myr})$.

We also include stellar feedback from Type II supernovae, which deposits a fraction of the star's mass and energy back into the 
fluid quantities of the occupied cell over a dynamical time. The prescription is given by
\begin{eqnarray}
\Delta M_{\rm gas}	&=& f_\star m_\star \\
\Delta E_{\rm gas}	&=&	(1 - f_{\rm CR}) \epsilon_{\rm SN} m_\star c^2 \\
\Delta E_{\rm CR }	&=&	f_{\rm CR} \epsilon_{\rm SN} m_\star c^2 \; ,
\end{eqnarray}
where $f_\star = 0.25$ is the mass fraction of the star ejected as winds and SN ejecta, $\epsilon_{\rm SN}$ is the Type II supernovae efficiency and $f_{\rm CR}$ is the fraction of this energy feedback donated to the relativistic CR fluid. For our fiducial run we set $\epsilon_{\rm SN} = 10^{-5}$, corresponding to $10^{51}$ ergs for every $55 M_\odot$ of stars formed. We also typically set $f_{\rm cr} = 0.3$, in line with current observations \citep{Wefel1987}.

\section{Results}
\label{sec:results}

We now present the results of our simulations, first describing the outcome of our run with all parameters set to their fiducial values, and then exploring how the results depend on the parameter values.  This will allow us to gauge how robust our results are to small changes in the model, and will help us gain intuition into what role the various physical processes play. A majority of the analysis presented here was facilitated by the simulation data analysis and visualization tool {\tt yt} described in \cite{Turk2011}.

Table \ref{parameters} provides a description of the parameters varied in this work. The central row describes all parameter choices for our \emph{fiducial} run. The other entries of the table represent single-parameter deviations from the fiducial case in our 20 additional simulations. Entries in resolution columns denoted with * were run both with and without the CR fluid. The varied parameters are (from left to right) spatial resolution, $\Delta x_{\rm min}$, in parsecs; mass resolution, or resolution of the base grid ``size''; maximum CR 
sound speed, $c_{s,{\rm max}}$, in km/s (see Section~\ref{sec:implementation}); star formation efficiency, $\epsilon_{\rm SF}$; supernova feedback efficiency, 
$\epsilon_{\rm SN}$; the fraction of energy feedback diverted into the CR fluid, $f_{\rm CR}$; the CR diffusion constant, 
$\kappa_{\rm CR}$ in cm$^2$/s; and the power index for the CR equation of state, $\gamma_{\rm CR}$.

\subsection{Fiducial Run}

We now describe the results of our fiducial run, which has CR diffusion and parameters set to observationally or physically motivated values.  We will begin with a visual examination of the results, before moving on to 1D profiles and finally the evolution of global values.

\subsubsection{Visual Evolution}

\dubFig{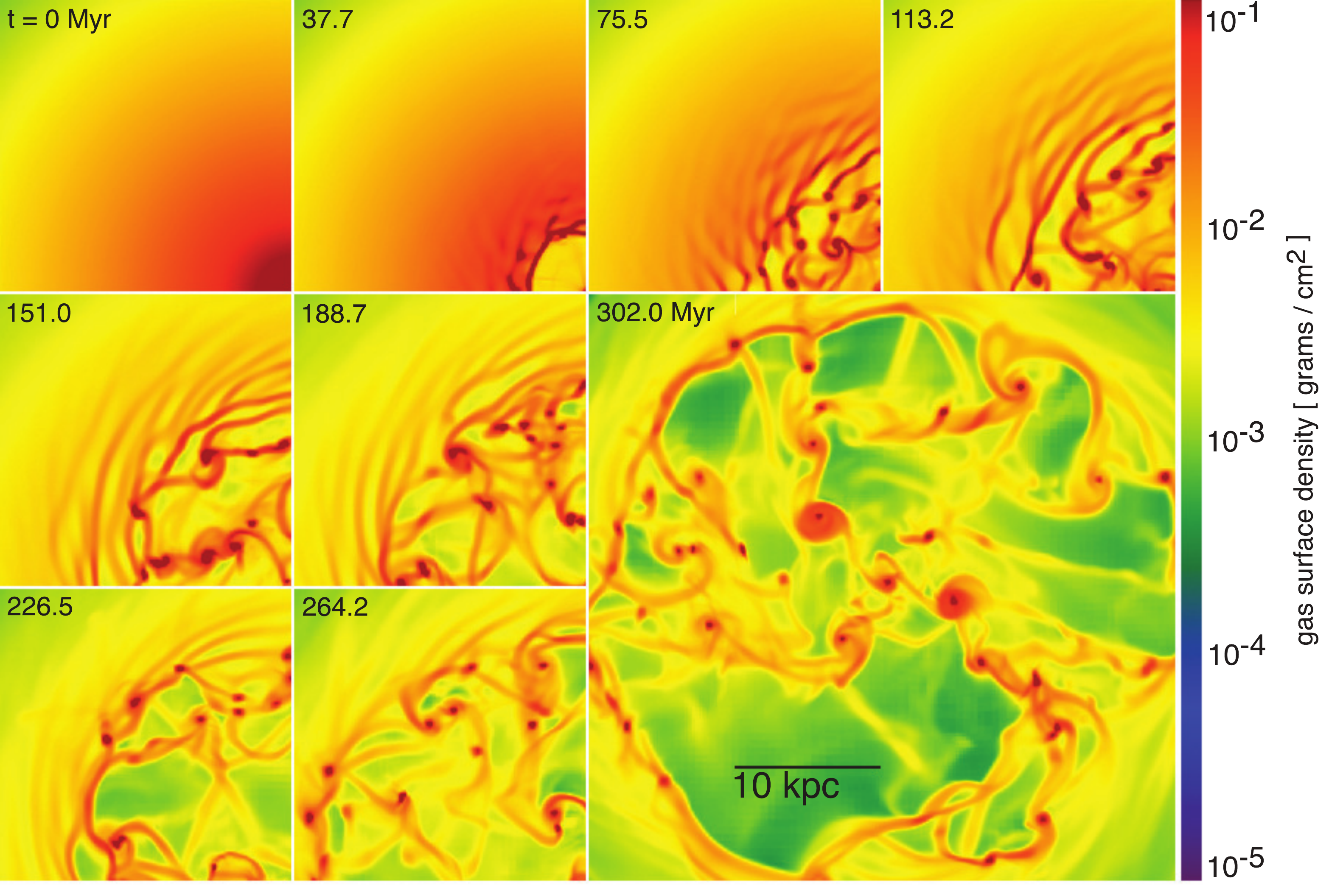}{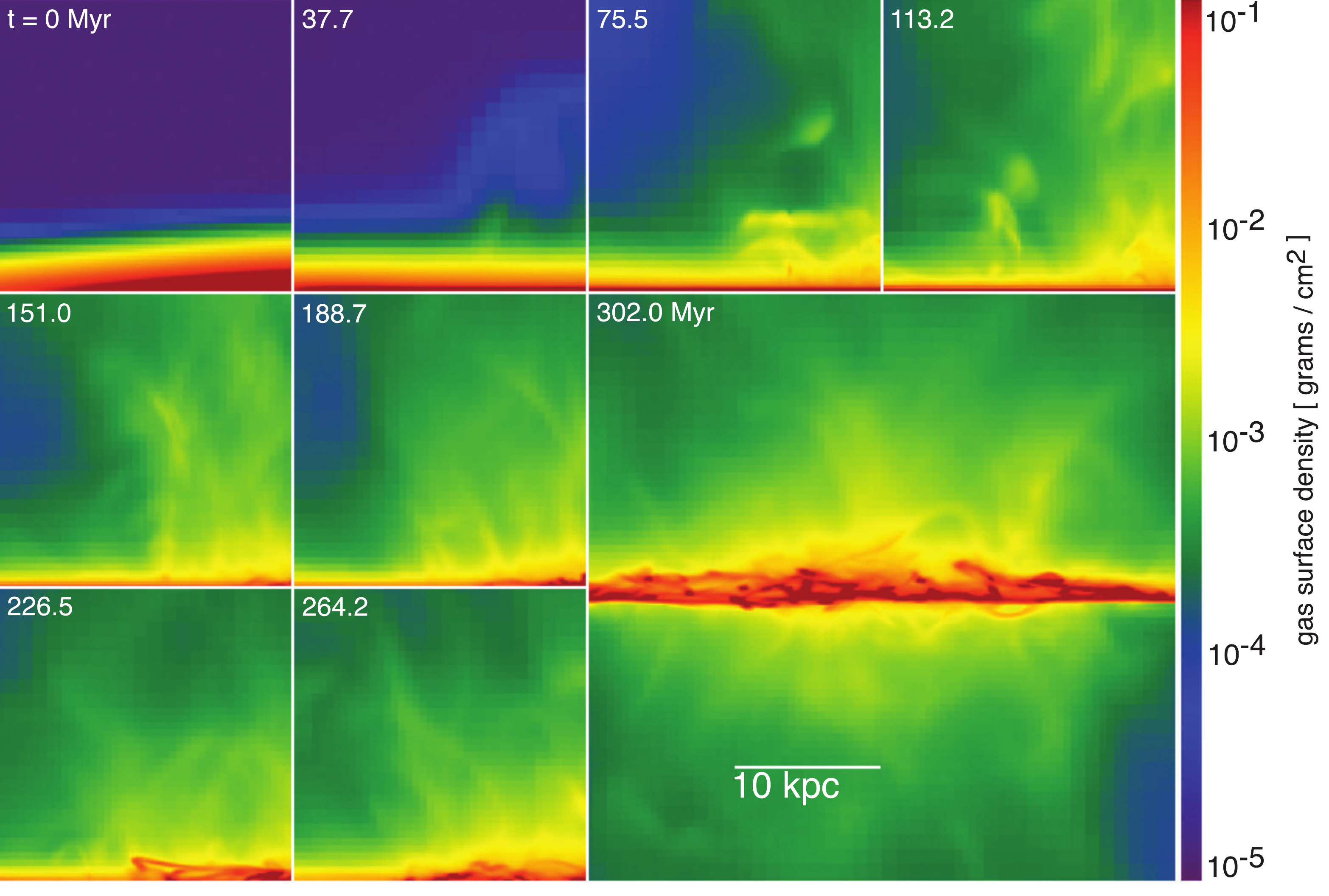}
	{Face on (top) and edge on (bottom) projection of gas density for our fiducial run at various times, as indicated. In all but on panel, we show only a quarter of our full disk.}{density}{.44} 

The top half of Figure \ref{density} shows a ``face-on'' view of the gas surface density as our run progresses. Though somewhat 
altered by the presence of CRs, the disk evolves in a manner quite similar to that described in \cite{Tasker2006}: it cools down to 300K, and
the gas quickly slims to less than a kiloparsec in thickness, beginning in the galactic center where the dynamical time is smallest. 
The collapse then ripples outward as spiral filaments funnel gas into knots. These knots exceed their surroundings in 
density by over two orders of magnitude, and act like softened point-sources of gravity, scattering off one another and making 
small excursions from the disk. At late times their number and size stabilizes within the unstable portion of the disk.

The bottom half of Figure \ref{density} shows an ``edge-on'' view of gas surface density, where immediately evident are robust, 
filamentary flows of gas out of the disk and into the galaxy's halo, beginning about 50 Myr after the start of the simulation (coincident with the start of a strong starburst, which will be discussed in more detail below).  
In the innermost regions of the halo, the highest surface 
densities can be found just above where the collapse and fragmentation of the galactic disk proceeds radially outward. However, the 
projected density appears more homogenous far from the disk, especially at later times, where it fills spherical lobes above and 
below the plane of the disk with densities around $10^{-26}$ g cm$^{-3}$. These lobes grow continuously, meeting the boundary of 
our 500 kpc-cubed simulation box by roughly 500 Myr, implying an average speed for the shock front of $\approx500$ km~s$^{-1}$.

\dubFig{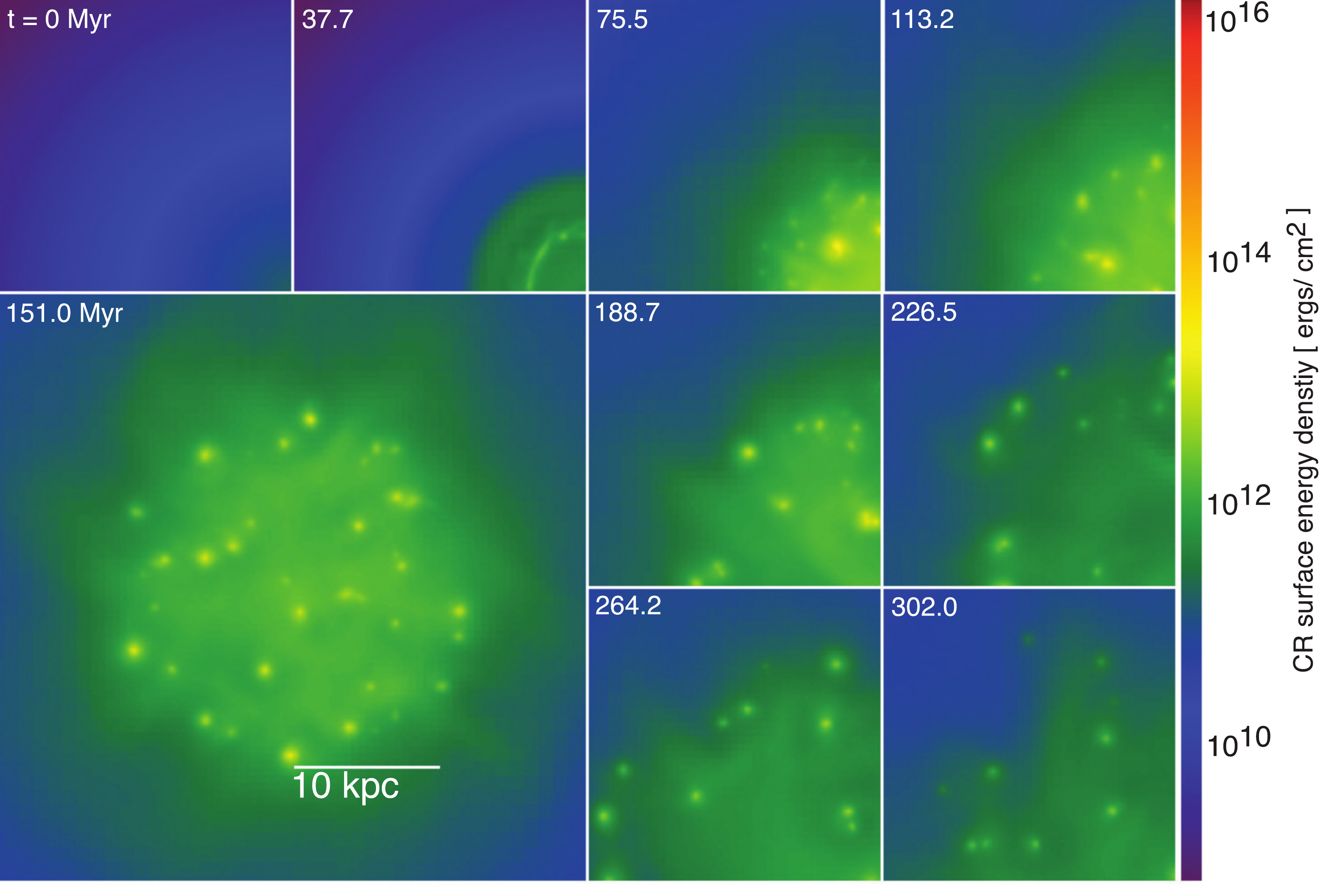}{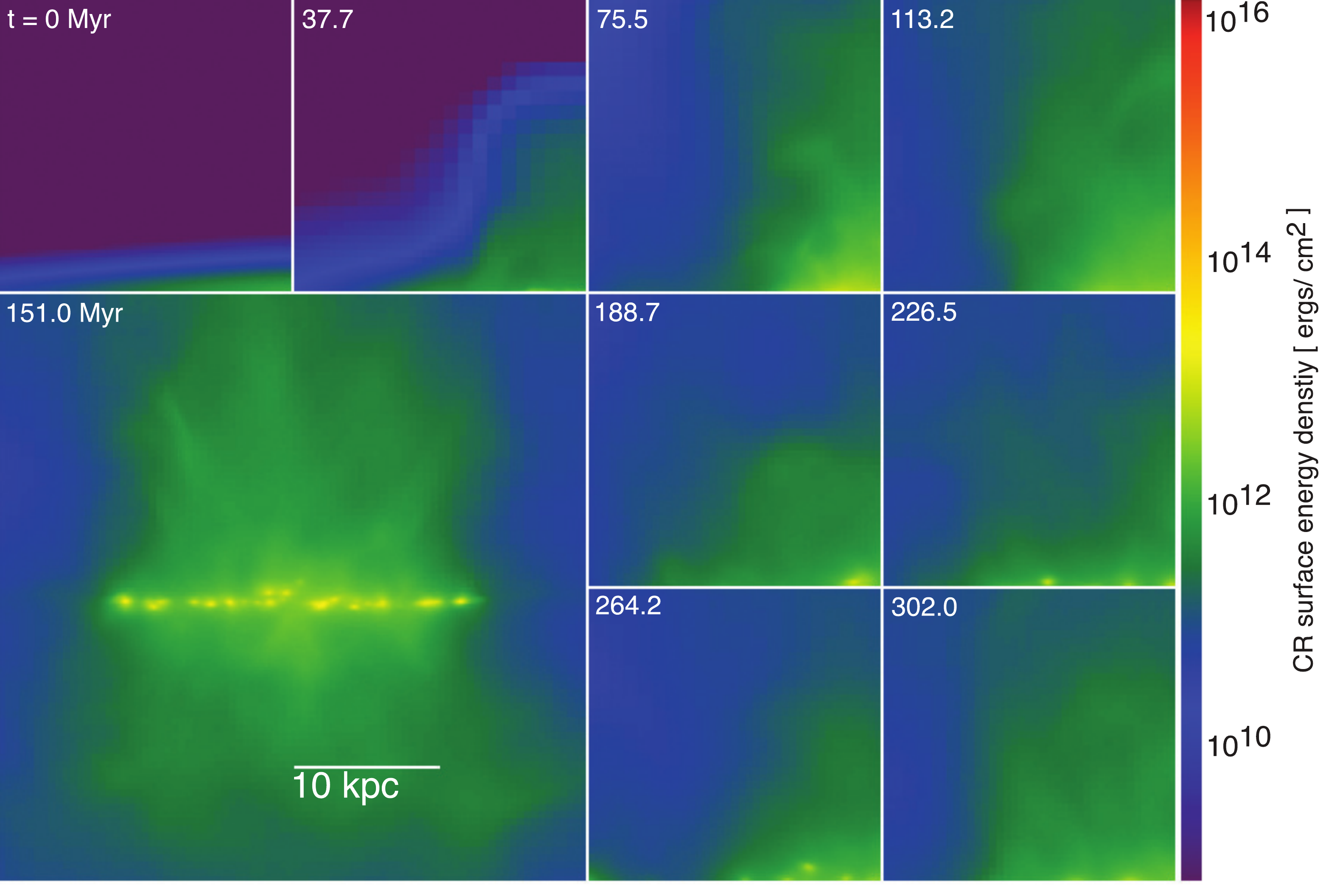}
	{Face on (top) and edge on (bottom) projection of CR energy density for our fiducial run.  In all but the final panel, we show just one quadrant of the disk.}{crs}{.44}

Figure~\ref{crs} shows projections through the simulation box of CR energy density, which we will refer to as CR surface energy 
density, or simply CR surface density. Recall $\epsilon_{\rm CR} = ( 1 - \gamma_{\rm CR} ) P_{\rm CR}$, and thus we can regard 
bright regions of these plots as areas of high CR pressure, where the rays may work to liberate gas from bound structures. The 
rays initially populate the simulation box in a profile identical to the thermal gas; however the CR distribution is quickly dominated by the ongoing injection from star formation.  As described earlier, we add CRs into cells where stars form, and thus the 
face-on projections show bright clumps that trace recent star formation. These clumps coincide with the dense knots in the thermal 
gas plots, home to the most vigorous star formation. The rays advect along with the thermal gas, but unlike the thermal gas the 
rays are also highly diffusive; over time these clumps dissolve to fill their surrounding regions, both within and above the disk. From 
the edge-on view in Figure \ref{crs}, we see rays flow out into the halo in a manner similar to the thermal gas, but with a smoother distribution.  

\stdFullFig{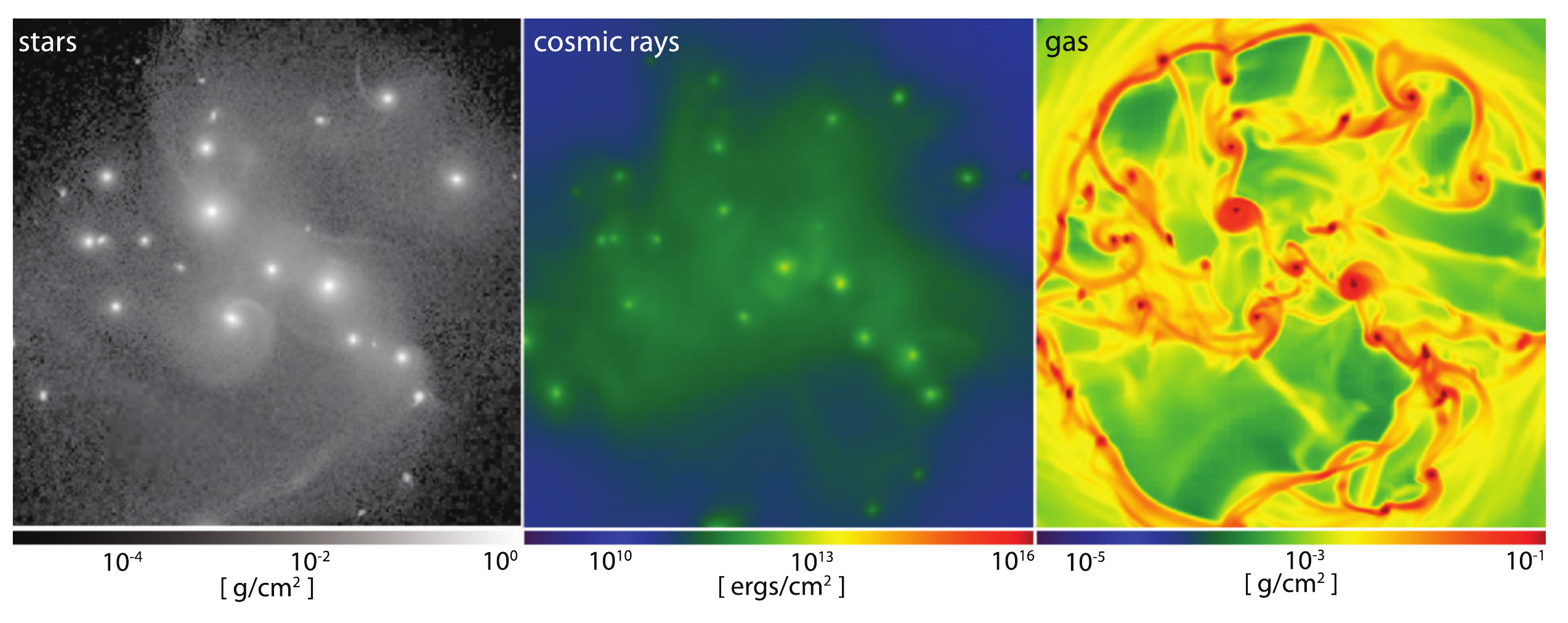}{The surface density of stars (left), cosmic rays (center), and gas (right) at $t=302$ Myr. Although there exists a 1-to-1 correspondence
between clumps in all three quantities, many of the brightest star clusters are much fainter in CR surface density, implying that these clumps are older, and producing
fewer new stars (and thus fewer CRs). The diffusive CRs show the least structure, instead tracing very recent star formation.}{star-comp}{.65}

Figure \ref{star-comp} shows a face-on projection of stars in the disk compared to both CRs and gas density. Bright clumps of 
stars and CRs show a one-to-one correspondence across the projections, although some of the largest central star clumps do not
figure as prominently in the CR surface density because they are older and so generating few CRs. The CR energy density in a clump is set by a competition between injection from star formation and diffusion/advection.  This results in the CR energy density having a lower contrast than the gas; however, a net pressure gradient in the CR component still persists, which -- as we will show -- can drive significant outflows.
Many star clusters have interacted, producing tidal tails. These gravitational features
are absent from the CR plots. These dense clumps also appear in the thermal gas plot (rightmost panel), although this fluid shows far
more filamentary/cavity structure than either the stellar or CR distributions. The CR fluid thus appears to be a good tracer of recent
star formation.

\stdFullFig{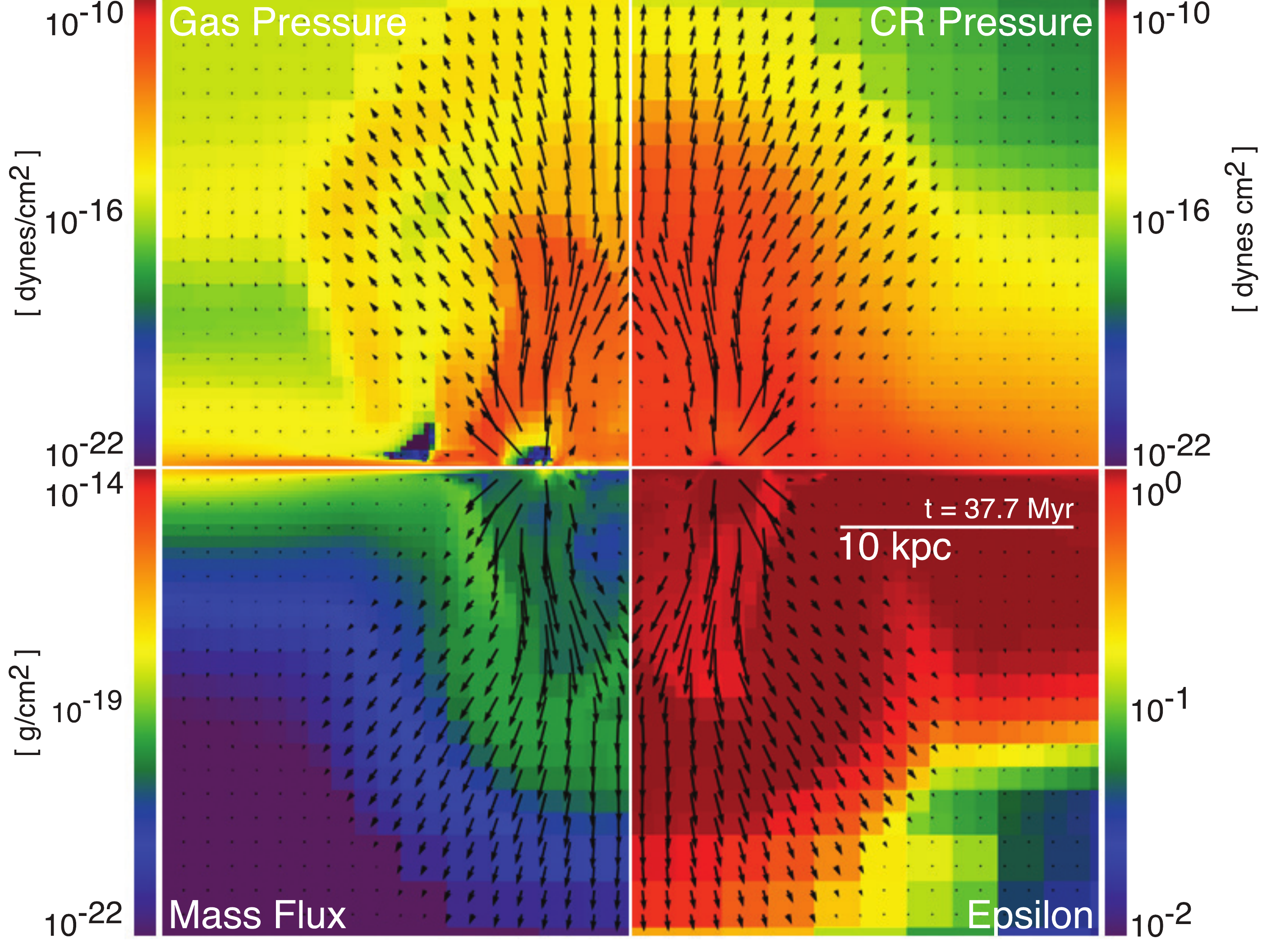}{Slices of mass flux, thermal gas pressure, CR pressure and $\epsilon = P_{\rm CR}/P_{\rm T}$ at $t = 37.7$ Myr during our fiducial run. 
This snapshot displays the most violent burst of star formation in the fiducial run, and thus an ideal study of the anatomy of our winds.}{pressure-flux-slices}{.5}

We can better understand these flows by plotting mass flux and both relevant pressures (thermal and CR). Figure~\ref{pressure-flux-slices} does so 
at $t=37.7$ Myr, during an early burst of particularly intense star formation. Here we show an edge-on slice through the galaxy, in four 
different quantities. Since these flows exhibit noticeable asymmetries, Figure~\ref{pressure-flux-slices} shows only the 
upper right quadrant of the slice in each quantity, flipped horizontally and vertically to appear as a complete picture. An indicated in the figure caption, the quadrants represent: 1) pressure of the thermal gas, 2) pressure of the cosmic ray fluid, 
3) vertical mass flux, defined as $\rho | \textbf{v}_z |$, where $\hat{z}$ is perpendicular to the galactic plane; and 4) a ratio of 
CR pressure to combined pressure, $\epsilon \equiv P_{\rm CR} / \left( P_{\rm th} + P_{\rm CR} \right)$.  In this last quadrant, 
deep red implies strongly CR-dominated dynamics while blue implies strongly gas dominated. From these plots we see the gas 
accelerates close to the disk itself, in pockets of strong CR pressure mostly devoid of gas pressure. Ahead of these fast, 
evacuated flows is a denser, slower component that carries more mass flux. Beyond the current reach of the diffusive rays the 
halo sits dormant, in hydrostatic equilibrium. 

Figure \ref{pressure-flux-slices} can be regarded as a caricature of the run at large: At $t=37.7$ Myr the initial conditions are collapsing into a cooler 
disk, and the SFR is $\sim 400$ M$_\odot$ yr$^{-1}$, a very large burst. Much later in the run, the SFR has fallen to $\sim 50$ M$_\odot$ yr$^{-1}$, and the 
acceleration of gas out of the plane has likewise fallen, but the qualitative features of this scene persist, and mass flux falls off 
less rapidly than the SFR. At later times, the acceleration region (where CRs dominate the pressure) has grown tremendously , providing a gentler acceleration that nevertheless persists to high altitude, which we describe in a more quantitive fashion next.

\subsubsection{1-Dimensional Profiles}

\stdFullFig{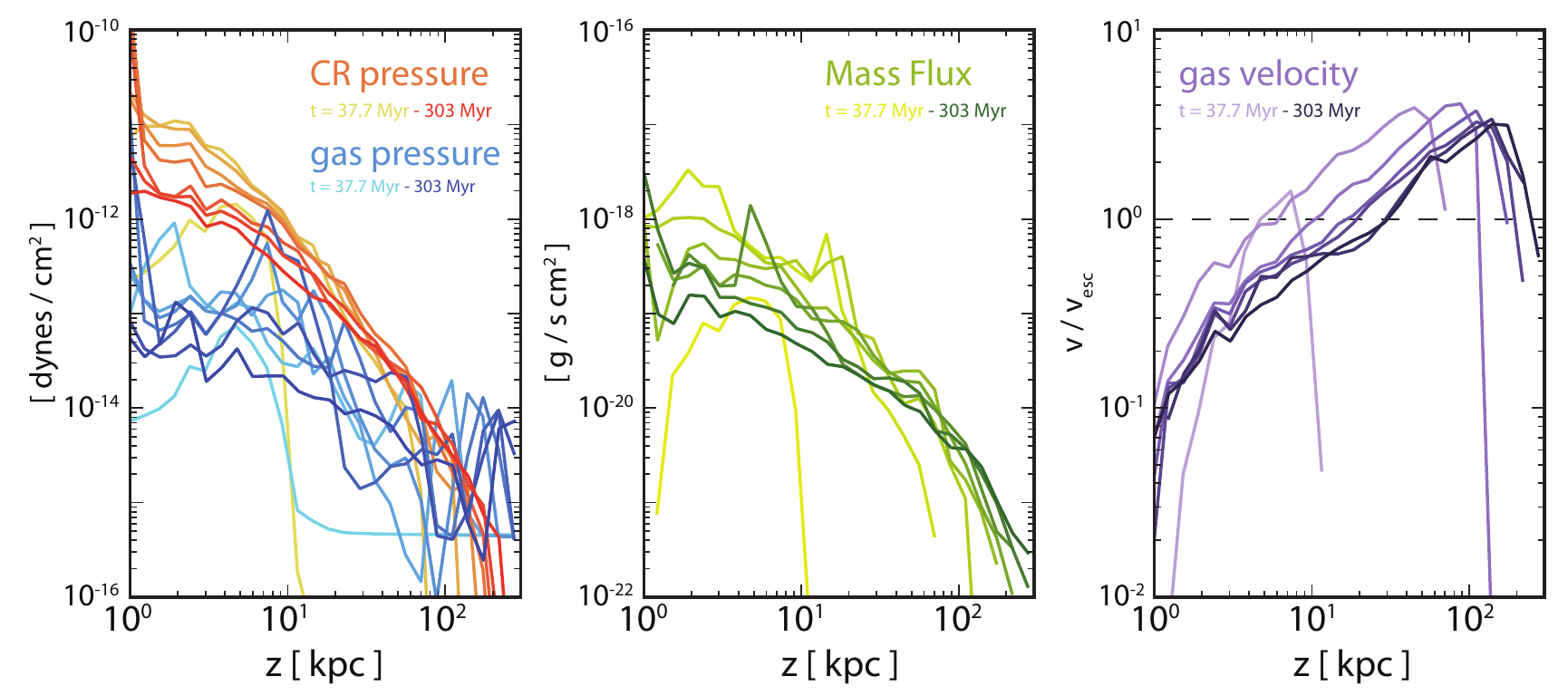}
	{Profiles of fluid quantities as a function of height above the disk plane, $z$. Here we ``bin'' the data, averaging over all cells inside a cylinder of radius 50 kpc at a given height in a mass-weighted fashion. Lighter colors represent earlier times, plotted in $\sim38$ Myr increments for roughly 300 Myr. The leftmost panel plots pressure of the thermal gas (blue) and cosmic rays (orange). The central panel plots vertical mass flux away from the disk, $\rho \textbf{v} \cdot \hat{z}(z/|z|)$. The final panel plots the ratio of gas velocity to the escape velocity at that height (again, a mass-weighted average of all gas at a given height above the plane), where a value of unity (indicated by a dashed line) implies this gas parcel would escape the galaxy's halo, barring any subsequent hydronamic interactions. }{pressure-flux-profiles}{1}
	
To better understand the dominant role CRs play in these flows, we turn to 1D profiles of key quantities, shown in Figure~\ref{pressure-flux-profiles}. Here we plot the time evolution as a color gradient, over 300 Myr in $\sim40$ Myr intervals, with lighter colors representing earlier times. For these plots, we construct a cylinder of radius 50 kpc, centered on the galactic center, aligned with the galaxy's angular momentum vector. We then average the quantity of interest in a mass-weighted sense at a given height above the plane within this cylinder. Thus a data point on these plots represents an average of the quantity of interest within a wide disk, a distance $z$ from the galactic mid plane (both above and below), weighted by density to better reflect the state/dynamics of denser gas pockets.

The leftmost panel of Figure \ref{pressure-flux-profiles} plots both CR and gas pressures over our 300 Myr period of interest. Although our initial conditions place CRs in a secondary role throughout the simulation domain, they rapidly assert themselves as the dominant pressure source beyond the disk. As the winds push outward, to hundreds of kpc in height above the disk, the rays continue to dominate the dynamics, except in a swept up shell of gas at the forefront of the flow, where thermal gas pressure spikes. The slope of this pressure profile beyond 20 kpc goes roughly as $z^{-4}$, consistent with adiabatic expansion of our $\gamma = 4/3$ ultra-relativistic CR fluid for spherical outflows.

The central panel of Figure \ref{pressure-flux-profiles} plots vertical mass flux away from the disk: $\rho \textbf{v} \cdot \hat{z}(z/|z|)$. Close to the disk (within $\sim50$ kpc) this quantity falls as $z^{-1/2}$, suggesting the flow is rather collimated and the majority of mass continues to rise once it has left the disk. Its normalization rises rapidly at early times (consistent with the peak in the star formation rate, described below), before falling at late times, as star formation declines. Far from the disk, the flux drops off as $z^{-3}$, consistent with a constant spherical outflow. These profiles appear to rule out a primarily fountain-like flow of even our densest halo gas (recall these profiles are mass-weighted), at least in this extreme starburst setting.

In the rightmost panel of Figure \ref{pressure-flux-profiles}, we plot the mass-weighted average vertical velocity, $v_z$, of the gas, normalized by the escape velocity at that height. Gas above the dotted line, if free to follow a ballistic trajectory, would leave our $10^{12} M_\odot$ halo. In contrast to standard energy- and momentum-driven feedback models, our outflowing gas does not obtain it's full velocity near the disk -- instead the acceleration process is smoother, with CRs in the halo powering flows with increasing velocity tens of kpc above the disk. This gentle mechanism shows no sign of abatement at late times, even as the star formation rate has fallen far below the exotic ULIRG values of our early evolution.

\subsubsection{Global Quantities}
\stdFullFig{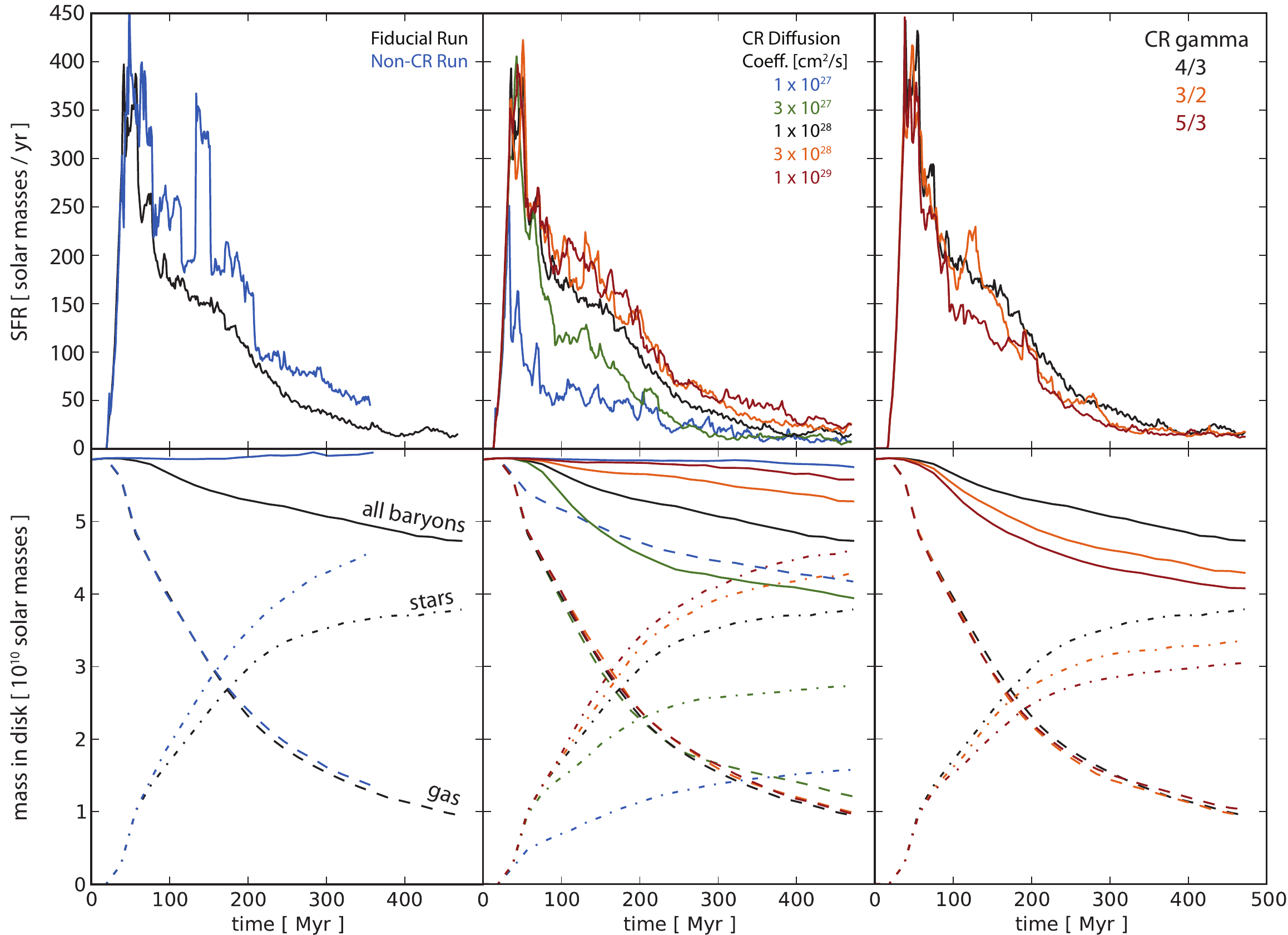}{Top Row: Star-formation rate (SFR) as our run progresses. Bottom Row: Total baryonic mass within the galactic disk over time, also broken down into stars and gas. The left-most column compares our fiducial run, described in the central row of Table \ref{tab:params}, against an identical run devoid of cosmic rays. The central column compares runs where we've varied the CR diffusion coefficient, $\kappa_{\rm CR}$. Finally, the right-most column compares runs with various $\gamma_{\rm CR}$.}{fiducial-global}{.7}

The outflows observed in these CR galaxy simulations should have meaningful implications for the global properties of our galaxy. The top-left panel of Figure~\ref{fiducial-global} shows the star formation rate for two runs as a function of time -- one is our fiducial run (with CRs and diffusion), and the other a run without any CR component.  Both simulations show an immediate burst of star formation, but the run with CRs has a lower SFR at almost all times.  Here we emphasize that both runs contain an equal amount of energetic feedback from supernovae: in the CR run, $30\%$ is injected into the CR fluid, whereas in the non-CR run, this energy is injected into the thermal gas.

The key result of these simulations lies in the bottom-left hand panel of Figure~\ref{fiducial-global}, where we plot the amount of gas and stars in the disk. Within 500 Myr, our CR-laced disk loses roughly 20 percent of its baryonic mass while converting roughly two-thirds of its gas to stars --- a mass loading factor of $\sim0.3$. No equivalent mass loss occurs in the non-CR run despite its supernovae feedback\footnote{The mass within our cylinder increases slightly in the no-CR run since radiative cooling allows outlying disk gas to come within 5 kpc of the central plane.}. This result is important, as \citet{Tasker2006} demonstrated that, regardless of parameter choice, it is very difficult to generate significant outflows with purely thermal feedback.  We will next explore how this result changes when we modify our numerical and physical parameters.

\subsection{Impact of the CR Diffusion Coefficient}
To better understand what role our choice of diffusion coefficient $\kappa_{\rm CR}$ plays in these dynamics, we ran three additional runs with higher and lower diffusion coefficients, one below the fiducial value of $10^{28}$ cm$^2$s$^{-1}$ and two above, in half-decade increments. We also performed a run with CRs but no diffusion. The central column of Figure~\ref{fiducial-global} shows the SFR and the baryonic mass in the disk over time for these runs.

We begin by looking at the case with no diffusion, described by the blue lines in both panels of Figure~\ref{fiducial-global}, an obviously unrealistic scenario that nonetheless provides insight into the mechanism behind our outflows.  In this case, the rays are completely tied to the thermal fluid, and the combined two-fluid acts almost like an adiabatic gas (since there is no cooling of the CRs).  This strongly suppresses the star formation but does not lead to any significant gas outflows. From the top-central panel of the figure, we see star-formation drops by roughly a factor of four compared to the non-CR run. Despite this effective feedback, the CRs cannot drive flows since supernovae deposit them only into the densest regions of the disk where star formation occurs. Bound to this gas, they remain dynamically subdominant. This leads to a thickening of the disk in what is essentially a convective process. And since the rays cannot diffuse beyond these dense regions, they cannot assert their presence in the lower-density regions of the disk where they effectively drive outflows in the runs with diffusion.

A remarkable thing happens when we turn on CR diffusion. For our lowest diffusion coefficient run ($\kappa_{cr} = 3 \times 10^{27}$ cm$^2$ s$^{-1}$), the SFR rises somewhat compared to the no-diffusion case, but the bottom-central panel of Figure~\ref{fiducial-global} reveals a qualitative shift in the dynamics: we immediately see strong outflows, leading to a mass-loading factor of nearly unity.  We therefore conclude that diffusion plays a crucial role in this process, moving CRs from pockets of very high density, where star formation occurs, to areas of the disk with lower density. Once out in the diffuse ISM, the rays dominate the dynamics, and the gradient in the CR fluid pressure works to accelerate disk material away from dense clumps and ultimately beyond the disk. We will discuss a simple model which captures these dynamics below.

However, if this diffusion occurs too quickly, the rays do not linger long enough to accelerate much gas: from Figure~\ref{fiducial-global}, we see the mass loading factor drops steadily as $\kappa_{\rm CR}$ rises. Thus the shorter the mean free path for the rays, the more important this mechanism will prove in the disk's evolution. As we increase the CR diffusion coefficient further, the SFR increases, approaching an evolution very similar to that seen in the case of no CRs.  As $\kappa_{\rm CR}$ rises beyond $10^{29} \; {\rm cm^2/s}$, the SFR approaches the no-CR case and the mass-loading drops towards zero. This picture is consistent with the na\"ive expectation that for very high diffusion coefficients CR-enhanced regions rapidly wash out, eliminating any CR pressure gradients and rendering the rays dynamically irrelevant.

\subsection{Impact of $\gamma_{\rm CR}$}

Our CR model as implemented in this present work is too simplistic to capture many subtleties of a real population of galactic cosmic rays. We assume an ultra-relativistic gas of CRs, and thus our second fluid's equation of state has an index $\gamma_{\rm CR} = 4/3$. In reality, this index depends on the distribution of cosmic rays in momentum space. When lower momentum, moderately relativistic rays dominate the population, gamma rises towards that of a thermal gas (5/3), and the ray fluid exerts a stronger pressure response (see Section~\ref{sec:missingPhysics} for details). The right-most column of Figure~\ref{fiducial-global} explores how changes in this index affect our outflows. From the bottom-right panel, we find that as the CR fluid becomes less relativistic, the outflows strengthen, presumably since the CR fluid becomes ``stiffer'', responding to compression with a more dramatic rise in pressure for the same energy injection. Thus our simplistic model's choice of $\gamma_{\rm CR}$ seems unlikely to exaggerate CR-driven winds (but see the discussion for other caveats to these results).

\subsection{Impact of Feedback Prescription}

\stdFullFig{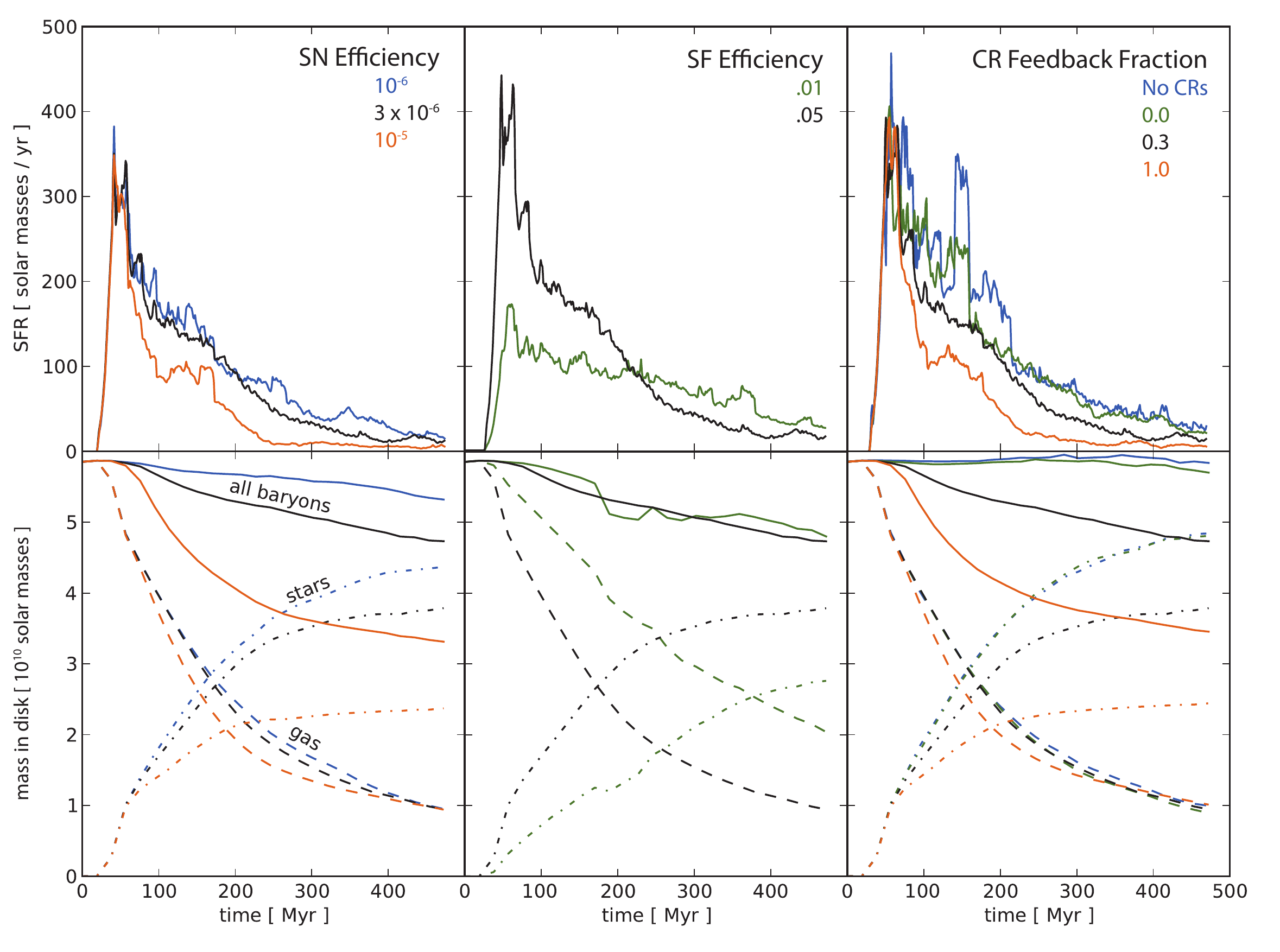}{The SFR rate (top row) and disk mass (bottom row) for a variety of runs which vary the SN Efficiency (left column), the SF efficiency (middle column), and the CR Feedback Fraction (right column). }{sn}{.7}

We continue our parameter study by investigating the role of our star formation and feedback parameter choices. Figure~\ref{sn} again plots the SFR and disk mass for these runs. In the leftmost column, we've varied $\epsilon_{\rm SN}$, the supernova efficiency, above and below the fiducial case by a half decade; the other parameters are left unchanged.  As demonstrated by the SFR plot, increasing the SN efficiency can suppress star formation by a noticeable fraction, though lowering the efficiency does not have as strong an effect. For all cases, the SFR tends towards a comparable, low, residual value at late times. The disk mass falls most dramatically for runs with high SN feedback efficiency: The traditional choice of $10^{-5}$ manages to liberate roughly half the disk mass with otherwise fiducial parameter choices. The mass-loading of our outflows seems to depend strongly on our choice of $\epsilon_{\rm SN}$. Although this parameter strongly affects how many stars we form and thus the gas fraction, as we saw earlier, it does not have a strong effect on the residual gas mass in our disk.

We next varied the star formation efficiency $\epsilon_{\rm SF}$, related to how much mass in a thermally unstable gas parcel is converted into stars. The SFR plot in the central column of Figure~\ref{sn} demonstrates that lowering this efficiency can strongly suppress star formation at the beginning of our run, when the rapidly cooling disk of pure gas first collapses. However, the SFR outpaces the fiducial run roughly halfway through the simulation, presumably as feedback becomes more important in regulating the dynamics. From the mass plot, we find the total mass ejected from the disk is roughly independent of this efficiency; however, the total residual mass of gas in the disk is approximately a factor of two larger for the low-efficiency case. This occurs because a lower efficiency means that gas must collapse to higher densities to match the same star formation rate as in the fiducial run (since the SFR $\propto \rho^{3/2}$).

We also investigated the role that $f_{\rm CR}$ --- the fraction of SN feedback given to the relativistic CR fluid --- plays in our model. The fiducial case sets $f_{\rm CR} = 0.3$, but we also investigated no CR feedback, $f_{\rm CR} = 0.0$, and complete CR feedback $f_{\rm CR} = 1.0$. As seen in the rightmost column of Figure~\ref{sn}, enhancing the fraction of feedback in the form of CRs allows them to more effectively suppress the SFR throughout our run. For the case where feedback is entirely thermal, the SFR is comparable to a run devoid of any cosmic rays. The mass plot shows that the mass-loading of CR-driven outflows strongly depends on $f_{\rm CR}$, with higher CR feedback liberating more disk mass.  A run with full CR-feedback and otherwise fiducial parameters can remove roughly half the disk mass.  On the other hand, when all of the energy is in thermal form ($f_{\rm CR} = 0$), no outflows are generated.  As in the case of $\epsilon_{\rm SF}$, the choice  of $f_{\rm CR}$ does not much affect the evolution of the residual gas mass in the disk. The $f_{\rm CR} = 0$ case features global SFR and disk mass evolution nearly identical to our non-CR run. This suggests our choice of CR initial conditions is unimportant, since CR diffusion quickly erases this information and the presence of CRs in the disk over long times is entirely regulated by star formation. From phase plots, we find the ratio of CR-to-gas energy density in the disk is consistent with observations in the solar neighborhood for runs with $f_{\rm CR} = 0.3$.

\subsection{Impact of Numerical Parameters}
\label{sec:numerics}
\stdFullFig{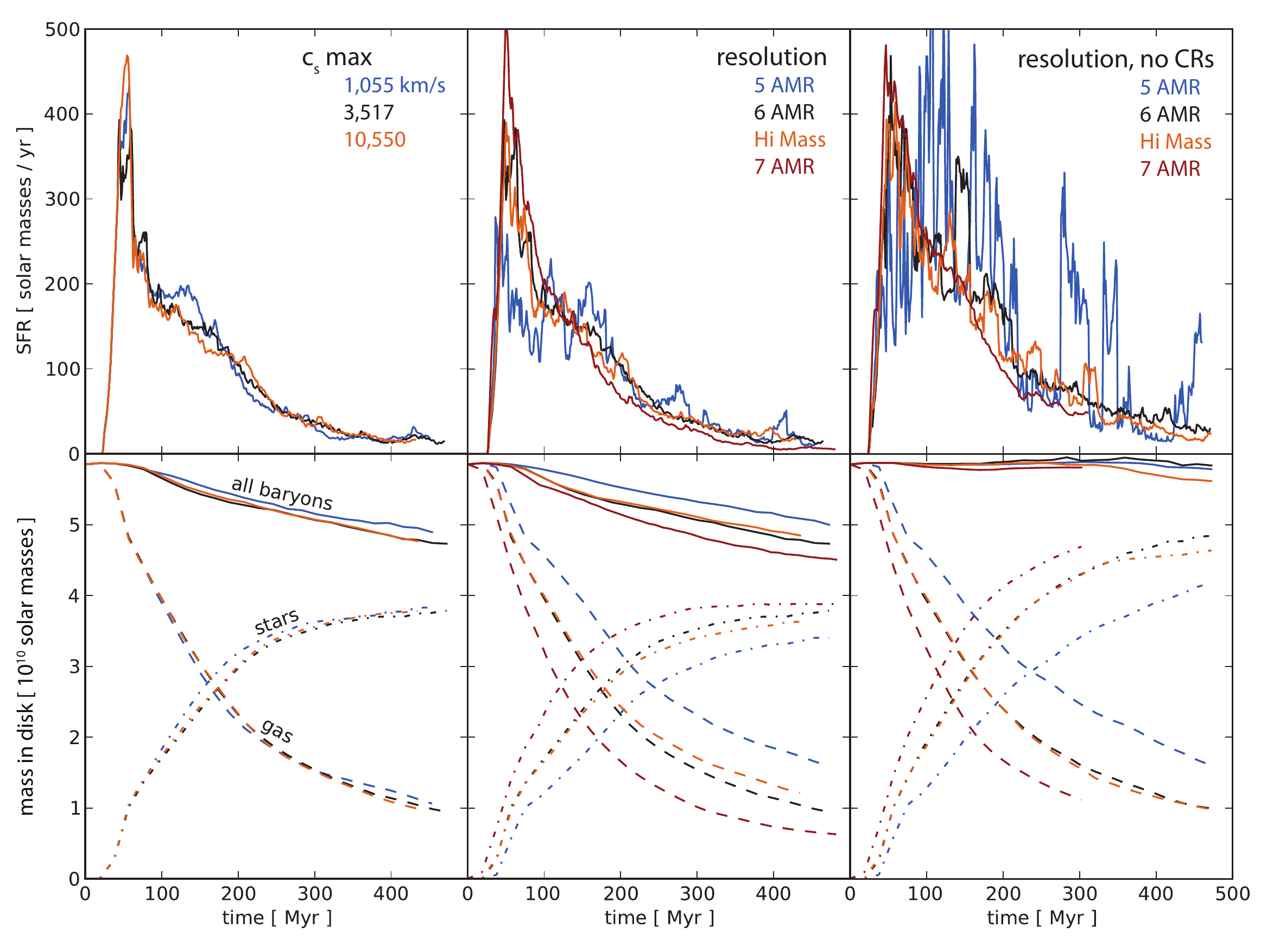}{The SFR rate (top row) and disk mass (bottom row) for a variety of runs which vary our numerical parameters.  In the left column, we vary $c_{\rm s,max}$, in the middle column we vary the maximum resolution by changing the maximum allowed level of refinement for runs with CRs, while in the right column we vary the resolution for runs without CRs.  In the resolution study, ``Hi Mass'' refers to a run with the same spatial resolution as the 6 AMR run, but an improved mass resolution (see text).}{numerical}{.7}

We wrap up our parameter study by exploring the impact of the primary numerical parameters important in these simulations.

As CRs diffuse into the galactic halo they can evacuate cavities near the disk which have low levels of thermal gas, and the cosmic-ray pressure can strongly dominate over the thermal pressure. This can be seen in Figure~\ref{pressure-flux-slices}, and tends to happen in the early evolution of the system. As discussed earlier, these high sound speed regions can dramatically slow the pace of our runs, and thus we elected to implement a maximum sound speed (via the gas density). For our fiducial run, we chose a $c_{\rm s, max} = 1,055$ km/s. From phase diagrams of density-vs-CRs, we can confirm a very small portion of the gas in our runs (by mass or by volume) feels the effect of this sound-speed ceiling. To confirm this somewhat arbitrary parameter does not alter our results, we also performed runs with $c_{\rm s, max} = 3,518$ and 10,550 km/s. The leftmost column of Figure \ref{numerical} demonstrates that the choice of $c_{\rm s, max}$ has little effect on our primary results, even demonstrating convergence as we raise the ceiling.

Finally we explore the role of resolution. Our fiducial run has a base-grid resolution of $128^3$ and six levels of AMR. For our 500 kpc$^3$ simulation box, this corresponds to a 61 pc maximum resolution in each dimension. We performed runs with 5 and 7 levels AMR, corresponding to 122 and 31 pc maximum spatial resolution, respectively. The choice of base grid sets the \emph{mass} resolution of our run. We thus also carried out a run with a $256^3$ base grid but only 5 levels of AMR, thus replicating the spatial resolution of the fiducial run but with eight times better mass resolution. The center column of Figure \ref{numerical} shows how these choices impact our runs. From the plots, we find that lowering the spatial resolution of our runs can work to suppress star formation, diminishing the role of CR-driven outflows and enhancing the final gas fraction of our disk. These relations simply reflect our star formation law, where the SFR scales with gas density to a power larger than one. Higher spatial resolution runs can resolve the collapse of cold gas to higher densities, and thus produce more rapid star formation \citep{Tasker2006}. CRs do not play a central role in this scaling of SFR with spatial resolution. To demonstrate this, we also performed our resolution runs without the  two-fluid model, shown in the rightmost column of Figure \ref{numerical}. The SFR and gas fraction scale over time in an identical fashion for these non-CR runs. With and without CRs, our choice of mass resolution has little effect on the plotted quantities.

\section{Discussion}
\label{sec:discussion}

\subsection{A Simple Model for CR-Diffusion-Driven Outflows}
\stdFig{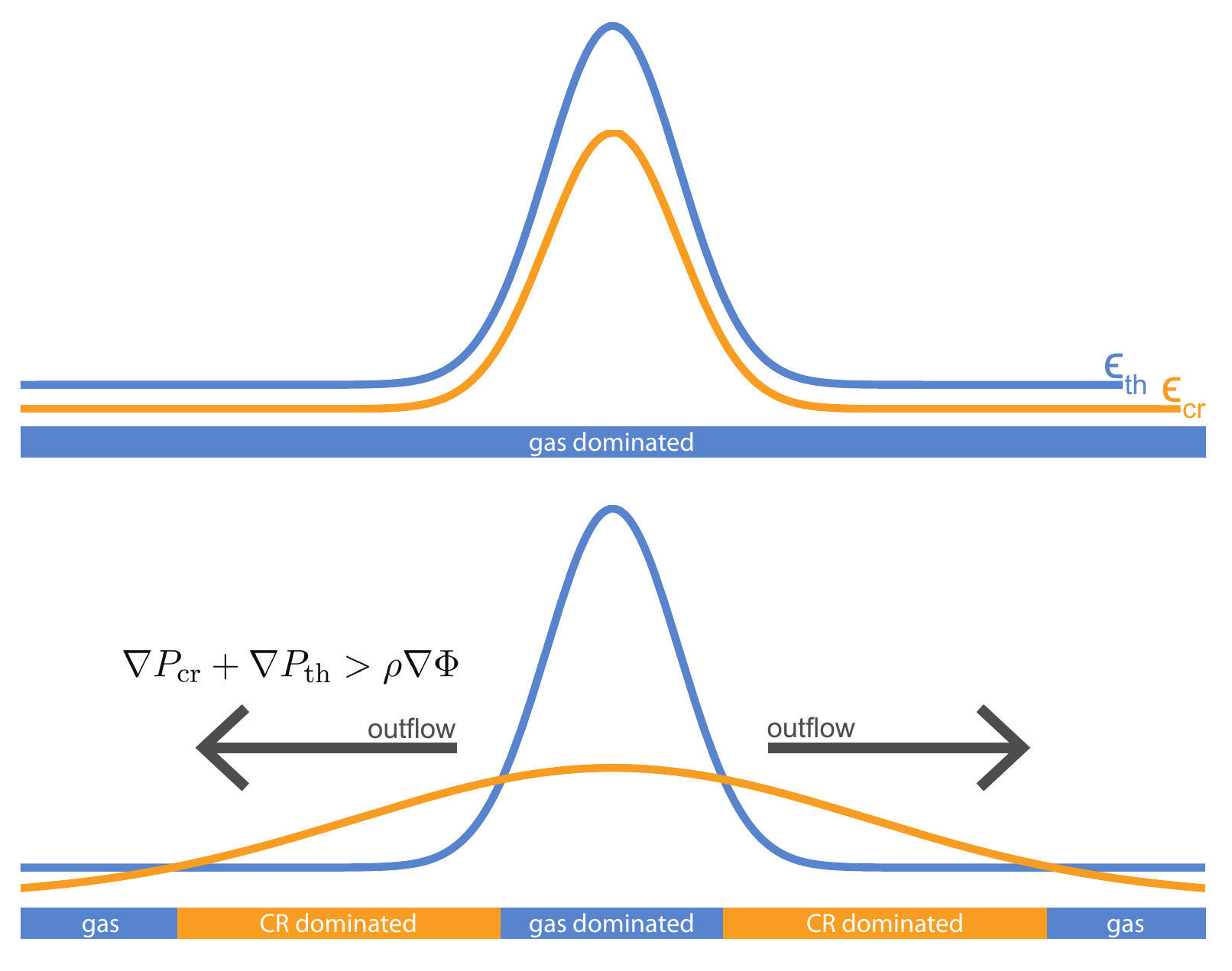}{A schematic model of CR diffusion-driven outflows.  The blue (gold) line shows a hypothetical gas (CR) density profile.  The top panel depicts a point early in the evolution, while the bottom panel shows a time after the CRs have begun to diffuse out of the clump. }{model}{.5}

As our parameter study demonstrates, including the cosmic ray fluid in our simulations will only launch mass-loaded, galaxy-scale outflows when the CR fluid is diffusive. The stronger the CR feedback and the longer CRs linger in the disk (i.e. the smaller the diffusion coefficient), the more gas can be ejected. This evidence points to a basic model for the outflows, illustrated in Figure~\ref{model}. The top panel of the figure illustrates a clump of dense gas, formed as the unstable portion of the disk begins to fragment, where both the matter and CR energy density become enhanced. The enhancement in CRs occurs for two, related reasons: the first is CR acceleration through SN resulting from star formation in the core, and the second is the compression of the CR fluid as the clump collapses gravitationally.  

At first, the CR fluid is everywhere secondary in strength to the thermal pressure, and does not dominate the dynamics. However, very quickly, the diffusive CR fluid begins to spread out, as illustrated in the bottom panel of Figure~\ref{model}, and its width grows beyond that of the thermal gas. Now the lower-density wings of the thermal gas feel the presence of an enhanced CR pressure gradient sloping outward, away from the clump's center. This pressure gradient exceeds the local self gravity of the gas and accelerates the double-fluid away from the clump. In a thin disk, the flow continues unimpeded out of the plane, and many clumps throughout the disk conspire to drive a galactic scale flow away from the mid-plane.

The timescale for CR diffusion for our choice of parameters is comparable to the dynamical time of the clump, and so these steps are not cleanly separated as shown in the cartoon. In our simulations, the rays actually provide an immediate form of feedback, even before star formation commences due to the compression of pre-existing cosmic rays. But without replenishment of the CR population via SN shock acceleration, the local CR peak can only persist a few Myr before diffusion and advection of the accelerated material depletes the CR reservoir. We find that star formation accelerated CRs are crucial for driving extended winds.

This model shows why diffusion is essential for driving outflows.  Without diffusion, the CR fluid never gains pressure dominance, particularly in the lower density regions of the cloud for which a given pressure gradient will cause a larger acceleration.  This explains why in simulations without diffusion the CRs act to puff up the disk but cannot drive significant outflows. \cite{Uhlig2012} came to a similar conclusion for their SPH runs with CR feedback, though they modeled CR streaming rather than CR diffusion (discussed below). They found that turning on this streaming, so rays could rapidly escape the densest star forming regions, was likewise necessary to drive outflows.

It also helps us to understand the observed dependence of outflow strength as we vary the physical parameters.  As we saw in the simulations, a lower diffusion coefficient leads to larger outflows.  In our model as depicted in Figure~\ref{model}, the timescale for spreading out of the CR profile is directly proportional to the diffusion coefficient, and if we make the simplifying assumption that the gas does not move significantly during this process, we see that a given parcel of low density gas in the wings will only feel the CR pressure gradient for this time period.  Therefore, the resulting velocity of the gas is proportional to $\kappa_{\rm CR}$, and more of the gas will exceed the escape speed, exactly as observed.

Most of the other parameters are even more straightforward -- a higher SN energy, or a higher CR fraction will result in larger pressure gradients for a given diffusion strength, and so stronger outflows.  The star formation efficiency is less obvious, although qualitatively we see that for a lower efficiency a given SFR (and likewise CR generation rate) will be delayed until the central clump density is higher; however, for the other parameters held constant, the CR acceleration is unaffected, as observed.

Finally, we note that the model indicates that the dense gas in the center is not accelerated by the CR fluid.  This is also observed in the simulations, with star forming clumps (molecular clouds) lasting for tens of Myr (or longer).  This indicates that CR feedback is not an efficient way to disperse molecular clouds, which is not surprising --- as we discuss in more detail below, another physical mechanism (e.g. radiation pressure, stellar winds) is required.  This also limits the amount of gas ejected since in our simulations the highest mass-loading we achieve (the ratio of mass ejected to mass of stars formed) is roughly unity.  But this is not a fundamental limit for this mechanism: higher mass-loading could be achieved if molecular clouds were dispersed into the ISM with another feedback mechanism.

\subsection{Implications of our Model}

The standard picture of SN-driven galactic-scale winds places the thermal gas in a starring role: the star forming region endows the gas with enough energy and/or momentum to rise out of the galaxy's potential while entraining the denser ambient medium in its path. In this model, diffuse, hard X-ray emitting gas at $\sim10^7$ K fills the majority of the volume, acting like a piston to sweep up a shell of denser gas~\citep{Chevalier1985,Strickland2000}. This dense forerunner may succumb to Rayleigh-Taylor instabilities and the hot wind escapes as denser clumps fall back towards the disk in a ballistic ``fountain'' fashion. As the evacuated gas mixes with cooler components of the halo and climbs beyond the galaxy's gravitational potential, it continually decelerates before it manages to escapes the galaxy, delivering heavier elements to the IGM~\citep{Strickland2007}. With reasonable parameters, high-resolution simulations of this model fail to launch appreciable mass into the IGM \citep[e.g.,][]{MacLow1999, Melioli2013}, although they may expel a significant amount of energy and metals.

The CR-diffusion driven winds we find here depart from this traditional picture in many fundamental ways. Here we identify two key differences with observational consequences.

\begin{enumerate}

\item Beyond our galaxy's disk plane CR pressure dominates the dynamics, launching slower, more massive winds where all but the densest clumps continue to accelerate throughout the flow. In stark contrast to a ballistic, momentum-driven feedback approach, our winds start slow, climbing towards the escape velocity $\sim10$ kpc away from the disk. Recent observations may favor this gentler, extended acceleration mechanism; \cite{Steidel2010} investigated the kinematics of 89 Lyman break galaxies with survey-quality far-UV spectra and found their features were well matched by a scenario in which the gas velocity \emph{rises} with distance, out to at least 100 kpc.

\item In our runs, while evacuated, $10^7$ K gas exists in pockets, particularly during the most extreme burst of star formation, the wind mostly comprises denser material below $10^6$ K, with an appreciable warm-ionized, $10^4$ K component. These winds deliver more disk mass to both the gaseous halo of the galaxy and the IGM beyond. Thus our model may help explain why star-forming galaxies show evidence for substantial amounts of multiphase gas in their halos \citep[e.g.,][]{Chen2010,Rubin2010,Tumlinson2011}.  

\end{enumerate}

\subsection{Missing Physics}
\label{sec:missingPhysics}
Although our simulations feature relatively high resolution, and by including cosmic rays we help push forward the science of modeling galaxies, we are well aware that our simulations are a cheap substitute for the turbulent, multiphase, magnetized ISM, rife with molecules, dust and radiation from massive stars. In this section, we briefly discuss many of the limitations of our work and even more briefly touch on their likely impacts.

We begin with resolution.  As in \cite{Tasker2006}, we find that the SFR throughout our run depends monotonically on resolution, with higher resolution runs producing more stars at a given time. Higher resolution runs can track collapsed gas into scales, and thus higher densities, where the SFR rises, as indicated by the Schmidt-type star formation law we adopt. However, these higher resolution runs may not produce more accurate SFRs in the disk, since at these smaller  scales feedback mechanisms we do not attempt to capture become important, as we noted earlier.  This makes it difficult to do a proper convergence study, although we did attempt it (see section~\ref{sec:numerics}).  However, it is likely that convergence will only come with a mechanism for dispersing molecular clouds.  Ironically, it may be that the lower resolution runs, which better match the Kennicutt relation, are more realistic models.  

Our simulations make no attempt at a self-consistent evolution for the magnetic fields. For this two-fluid picture to strictly hold as implemented, we require a stochastic, tangled field throughout our simulation  region. These inhomogeneities cause CRs to random walk through the fluid, thus obeying our simple model with advection and isotropic diffusion. Observations of both the Milky Way and other local galaxies indicate the magnetic fields within a galactic disk are roughly equally divided in energy between such a stochastic component and a large scale coherent field that traces the spiral structure~\citep[e.g.][]{Beck2013}. Thus CRs may preferentially stream within the plane of the disk, since the diffusion coefficient along fields lines is larger than perpendicular to them.  More work is required to better understand how diffusion depends on field topology, strength, and gas density \citep[e.g.,][]{Xu2013}; however, from our parameter study, we find our qualitative result does not change when we vary the diffusion coefficient by orders of magnitude and therefore we suspect that the basic picture of CR-driven winds does not depend strongly on how diffusion works.

We note that we also do not include the impact of cosmic ray streaming. In our model, the rays are tied to the field which is assumed to be frozen to the gas.  In reality, CR pressure gradients cause the rays to stream along field lines which can excite \Alfven waves leading to heating of the gas \citep{Skilling1975}.  This may play an important role several kiloparsecs from the disk, into the galaxy's halo, where some models expect a coherent magnetic field structure, with field lines rising perpendicular to the disk and leaving the galaxy in a radial fashion. As the CR fluid enters this regime, rays may no longer ricochet about the in-homogenous field and would instead stream along the coherent field lines at the \Alfven velocity. This would reduce the role of CR pressure in driving our outflows, but may imply a stronger role for an ``\Alfven pressure''  as described in \cite{Breitschwerdt1991}, although we note that CR pressure dominates in the 1D simulations of \citet{Dorfi2013}, which includes both components. Finally, within a kpc of the disk, at the disk-halo interface, the magnetic field structure is very uncertain. While CR streaming, \Alfven pressure and radiation pressure may prove more important for outflows in the outer regions of the galaxy, our results advocate an important role for CR diffusion-driven flows as the heavy lifters, accelerating appreciable disk mass out of the mid-plane star forming regions and into the halo. This highly mass-loaded flow in turn calls into question assumptions previous models have made about the field structure above the disk. If appreciable ISM rises into the halo, it may be that a more tangled, stochastic field might likewise get transported above the disk. A better treatment of MHD and star formation at higher resolution will be necessary to settle this issue.

Our CR fluid undergoes only adiabatic changes, except when bolstered by injections within star-forming regions and diminished by isotropic diffusion. In reality, diffuse shock acceleration on galactic scales and baryonic activity near the galaxy's supermassive black hole may also contribute to the CR fluid. In these non-cosmological runs, shock fronts do not play a central role in creating our galactic CRs. And for the purposes of this study, we wish to isolate star formation feedback from AGN feedback. Likewise, a more realistic model would capture CR loss processes, the most important being Coulomb losses and ``catastrophic losses''. In the former process, the charged rays slowly lose energy irreversibly to the surrounding plasma at large, heating the thermal gas while diminishing the CR energy density. The latter process involves the production of pions which decay into photons, electrons and neutrinos, resulting in a net loss of energy from the entire plasma via radiation. The relative importance of these two mechanisms depends on the distribution of CRs in energy: for CR fluids dominated by highly relativistic rays, catastrophic losses prove more important, and vice versa. Both processes scale inversely with the density of the thermal gas, $\rho$. An accurate calculation of these cooling rates also involves knowing the detailed momentum-space distribution of the CRs, since lower-momentum rays lose energy much faster than higher momentum, ultra-relativistic particles. Our present work makes no attempt at modeling the CR spectrum. In fact, we implicitly assumes the CR fluid is composed exclusively of ultra-relativsitic, $\gamma_{\rm CR} = 4/3$ rays, with a spectral index in momentum space of $\alpha = 2$. In their work, \cite{Jubelgas2008}, building on the work of \cite{Ensslin2007}, capture key aspects of the CR spectral distribution by assuming a CR distribution $d^2N/dPdV \propto p^{-\alpha}\theta(p-q)$ with constant spectral index $\alpha \in (2,3)$ and low-momentum cutoff $q$ (here $\theta$ is the Heaviside function). Within this framework, they calculate loss rates for the CR fluid, modeling the process as simply a rising cutoff, $q$, as low-$p$ rays lose their energy. They present cooling timescale curves as a function of the cutoff, $q$, for Coulomb and catastrophic losses. These timescales scale roughly with the inverse of gas density. If we assume a low-momentum cutoff of approximately $m_p c$, or higher, we find a lifetime for CRs of $\approx 1.2$ Myr in the densest star forming regions, $\approx .5$ Gyr in the disk at large, and  $\approx 10 - 1,000$ Gyr within our outflowing halo gas. A lower momentum cutoff can pull down these times scales an order of magnitude. Thus rays appear to be long lived, compared to the timescales of our dynamics, and thus our decision to ignore loss processes seems justified.

By forgoing a detailed description of the CR energy distribution, we also forfeit accurate knowledge of $\gamma_{\rm cr}$, a function primarily of spectral index, $\alpha$. Our fiducial choice of $\gamma_{\rm cr} = 4/3$ implicitly assumes $\alpha \to 2$, and thus the ``softest'' possible equation of state, where an adiabatic compression of the CR fluid produces a subtler rise in pressure than a thermal gas with $\gamma = 5/3$. Observations motivate a choice of $\alpha$ closer to $2.5$, and thus a somewhat stiffer CR fluid. Our parameter study has shown that a stiffer pressure response in the CR fluid enhances our outflows, so our fiducial runs are conservative in this regard.


\subsection{Comparisons with Previous Work}

Previous 1D models of CR-driven outflows have focused on diffuse winds ~\citep{Breitschwerdt1991,Breitschwerdt1993,Dorfi2013}. They take the disk-halo interface as the inner boundary conditions of the flow and assume straight, open magnetic field lines rising above the disk. Their runs include CR diffusion and streaming and an \Alfven wave pressure. The fast, diffuse flows of the standard wind picture presumably groomed the halo's magnetic field into this coherent structure, and the model thus appears self-consistent. Our results are broadly comparable with this work, particularly the time-dependent simulations of \citet{Dorfi2013}. In particular, they find that local CR enhancements close to the disk drive mass-loaded winds powered by CR pressure.

As described in the introduction, \cite{Ensslin2007} and \cite{Jubelgas2008} carried out simulations with the SPH code Gadget that included CRs using a somewhat more sophisticated model than this present work.  They found the CRs impacted the structure and star formation rate of their galaxies, particularly those with circular velocities below 80 km s$^{-1}$.  Most runs did not include CR diffusion, but they did include it for two runs of low-mass halos, where they found it significantly impacted the SFR.  It is unclear whether or not these runs featured significant winds, as found here.

More recently, \cite{Uhlig2012} simulated idealized galaxies in three-dimensions, including CR feedback using a modified version of the SPH code Gadget. Although they did not include diffusion, they implemented CR streaming, where rays flow down gradients in the CR energy density at speeds proportional to the local sound speed. Within this similar setting, they likewise found CR-driven outflows, albeit with some key differences. They found the inclusion of CR streaming crucial to this result for a similar reason as CR diffusion proves crucial to our present study: both mechanisms allow CRs to leave the densest star forming regions, where they are subdominant in the dynamics, into regions of lower gas density, where they can transfer energy via plasma waves and accelerate mass-loaded flows of gas. This streaming approximation makes strong assumptions regarding the magnetic fields: that at least a substantial minority of them have been combed open by supernova remnants, extending parallel to the gradient of CR energy density in the disk. Our model chooses to assume the opposite extreme: that field lines are tangled, and the liberation of CRs from the dense regions occurs via a random walk as they scatter off inhomogeneities in the field. \cite{Uhlig2012} found their outflow mechanism shuts off for massive halos, above $10^{11} M_\odot$. Their runs featured lower star formation rates, using a subgrid model better suited to quiescent galaxies with smooth, regulated SFRs \cite{Springel2003}.  In addition their runs begin with gas in a \emph{spherical} hydrostatic equilibrium. When they turn on radiative cooling, early outflows need to battle the ram pressure of inflowing gas raining down onto the disk. Our model instead begins with a more rotationally supported structure and a clear halo, which may better reflect the realities of a cosmological setting where gas streams in along defined filaments. Though they included CR cooling and loss mechanisms, their winds were not strongly regulated by these processes, suggesting our decision to ignore cooling is justified.

\section{Summary}

We performed the first three-dimensional, high resolution, adaptive mesh refinement simulations of an isolated starbursting galaxy that includes a basic model for the production, dynamics and feedback of galactic cosmic rays (CRs). This is one of the first 3D galactic disk simulations to include isotropic CR diffusion. We find CRs produced via supernovae-driven shock acceleration in star forming regions represent an important form of feedback, capable of suppressing star formation and driving mass-loaded, multiphase winds from a starburst galaxy within a $10^{12} M_\odot$ halo.

We implemented and tested a basic two-fluid model for the evolution of the thermal gas and the relativistic (CR) plasma, which captures the nonlinear interaction and evolution of these two components. We model additional relevant physics in our runs, including radiative cooling, shocks, self-gravity, star formation, supernovae feedback into both the thermal and CR gas, and isotropic CR diffusion, while we ignore other key components of realistic galaxies, including an explicit treatment of magnetic fields, CR streaming and loss processes, radiation pressure, stellar winds, and chemistry. Our galactic disk lies in a $10^{12} M_\odot$ halo within a $500$ kpc box, with adaptive resolution of up to $60$ pc.  

We ran a total of 21 simulations, exploring the consequences of various parameter choices related to the composition of our CR fluid, the details of our star formation algorithm and the key numerical parameters in our software, such as resolution. Below we summarize the key results of this work.

\begin{enumerate}

\item  The CR fluid is long lived and continually replenished during star formation, providing additional pressure support to the ISM and suppressing the global star formation rate (SFR).

\item A diffusive CR fluid can drive strong, massive, bi-polar outflows from a MW-sized ($10^{12} M_\odot$) starbursting galaxy, with a mass-loading factor of .3 ($\dot{M}/{\rm SFR} \approx 30\%$) for our fiducial case. For other reasonable parameter choices, the mass loading can exceed unity. 

\item We find that a mechanism such as CR diffusion (or possibly streaming) is crucial to this process.  Without diffusion, no wind is launched; however, as the diffusion coefficient decreases, the mass-loading factor of the wind increases, pointing to a picture in which diffusion moves CRs from high-density star forming regions to more diffuse areas of the disk where their pressure gradient can drive outflows.  Lower diffusion rates allow the CR pressure gradient to persist for longer, launching more massive winds.

\item These CR-driven outflows stand in contrast to thermal- and momentum-driven wind models, where hot gas ram pressure must rapidly entrain and accelerate the rest of the ISM. Instead, we see a massive, multiphase wind with slowly rising radial velocities over 10's of kpc. The relatively gentle acceleration results in a multiphase wind, which includes a cool, dense component that is generally not present in high-resolution thermally-driven winds.

\item The outflows strengthen when the star formation rate rises, the CR diffusion mean free path shrinks, or when a larger percentage of star formation feedback is apportioned to CR production. Although the relative strength of these outflows varies, their presence persists across wide swaths of parameter space, insensitive to the precise choices in our star formation model, the tuning of our CR fluid physics, the CR-diffusion mean free path and numerical parameters such as resolution.

\end{enumerate}

Our work suggests a new physical model for the generation of outflows from star forming galaxies. Although traditionally it has been argued that diffusion will lead to a homogeneous distribution of cosmic rays, we find that rapid star formation can maintain an enhanced CR presence at the disk mid plane capable of driving mass-loaded outflows by gently accelerating material and liberating appreciable gas from the halo. Thus cosmic rays may prove dynamically important to galaxy formation and evolution.

There are a number of enhancements to this work which should be investigated. One is to augment the CR physics captured by: (i) modeling the CR spectrum explicitly and thus allowing a basic treatment of radiative losses, and (ii) including MHD and anisotropic CR diffusion.  Another is to include the current CR model in cosmological simulations of galaxy formation, although it may prove challenging to match the resolution of the runs in this paper.  Finally, it would be interesting to investigate the observational implications of the simulations described here, to see how well the outflows match observed absorption-line studies of the circumgalactic medium.  Our work suggests that studying the detailed morphology of starburst superwinds can provide insight into the relative importance of various baryonic fluid components and the underlying structure of galactic magnetic fields, particularly at the disk-halo interface.

\section*{Acknowledgments}

We acknowledge financial support from NSF grants AST-0908390 and AST-1008134, and NASA grant NNX12AH41G,  as well as computational resources from NSF XSEDE, and Columbia University's Hotfoot cluster.

\bibliography{cr_paper}{}
\end{document}